\documentclass[fleqn,usenatbib]{mnras}

\usepackage{newtxtext,newtxmath}

\usepackage[T1]{fontenc}
\usepackage{booktabs} 
\usepackage{lscape}
\usepackage{array,longtable}
\usepackage{tabu}
\usepackage{hyperref}  
\usepackage{xcolor}

\DeclareRobustCommand{\VAN}[3]{#2}
\let\VANthebibliography\thebibliography
\def\thebibliography{\DeclareRobustCommand{\VAN}[3]{##3}\VANthebibliography}

\usepackage{graphicx}	
\usepackage{amsmath}	

\newcommand{\water}{H$_2$O} 
\newcommand{\methanol}{CH$_3$OH}

\title[Maser flaring in NGC6334I]{The Extraordinary Maser Flaring Event in the Massive Protostellar System NGC6334I: Multi-Epoch Milliarcsecond Resolution Investigation of the 6.7-GHz Methanol Masers}

\author[J. Kumar et al.]{
Jayender Kumar,$^{1,2}$\thanks{E-mail: jayender.kumar@csiro.au}
Simon P. Ellingsen,$^{2,3}$
Gabor Orosz,$^{4}$
Lucas Hyland,$^{2}$
Chris Phillips,$^{1}$
Cormac Reynolds,$^{1}$
\newauthor
Gordon MacLeod,$^{5}$
\\
$^{1}$CSIRO Space and Astronomy, ATNF Headquarters, 26 Pembroke Rd, Marsfield, NSW 2122, Australia\\
$^{2}$School of Natural Sciences, University of Tasmania, Private Bag 37, Hobart, Tasmania 7001, Australia\\
$^{3}$International Centre for Radio Astronomy Research, The University of Western Australia, 35 Stirling Highway, Crawley WA 6009, Australia\\
$^{4}$Joint Institute for VLBI ERIC, Oude Hoogeveensedijk 4, 7991PD Dwingeloo, The Netherlands\\
$^{5}$Hartebeesthoek Radio Astronomy Observatory, P.O. Box 443, Krugersdorp 1741, South Africa}

\date{Accepted XXX. Received YYY; in original form ZZZ}

\pubyear{\the\year{}}

\begin{document}
\label{firstpage}
\pagerange{\pageref{firstpage}--\pageref{lastpage}}
\maketitle

\begin{abstract}

We report the first multi-epoch milliarcsecond resolution imaging of the 6.7-GHz class~II methanol maser emission associated with the high-mass protocluster system NGC6334I. The observations covered a period of over 10 years in four epochs between March 2010 and March 2020. We confirmed for the first time the emergence of 6.7-GHz methanol maser emission associated with NGC6334I-MM1, which lies north of the previously known sites of class~II methanol masers, NGC6334-MM2 and MM3. The new maser emission was located close to the strongest (sub)millimetre source in NGC6334I-MM1, identified as MM1-B, which experienced a sudden increase in intensity in 2015, produced by an episodic accretion event. We are able to compare the location and intensity of the 6.7-GHz methanol maser emission among the epochs before, during, and after the flare, providing new insights into the relationship between maser flares and episodic accretion events in high-mass stars.

\end{abstract}




\section{Introduction}

The details of the processes by which high-mass stars form is a long-standing problem in astrophysics \citep[e.g.][]{MacLowKlessen:2004, KrumholzBook:2017, Meyer:2017}. The majority of high-mass star formation in the local universe takes place in clusters, but the detailed processes involved remain poorly understood. A range of theories have been proposed, but observational tests of them remain challenging as high-mass stars form in a protostellar cluster environment, deeply embedded within cold molecular gas and evolve on relatively short timescales \citep[e.g.][]{KrumholzBook:2017}. Episodic accretion is well established for young low-mass protostars \citep{Kenyon:1990, Hartmann:1996, Evans:2009, Moscadelli:2017}, but it has recently been identified towards a number of high-mass star-forming regions and appears to be a common aspect of the star formation process \citep{Meyer:2017}. Based on their hydrodynamic simulations, \citet{Meyer:2017} predicted that episodic accretion should also be present in primordial and high-mass star formation. The relatively small number of high-mass stars in the Milky Way means that episodic accretion events in this regime are more challenging in observational identification, as they will be less common, generally more distant and within a more complex environment. Because of the deeply embedded nature of high-mass star formation, one of the biggest challenges in identifying and studying episodic accretion is the timely detection of the onset of such events. 

Several studies have established an association between the flaring of interstellar masers and the accretion phenomenon \citep{Moscadelli:2017, Burns:2020}. It is now thought that monitoring of interstellar masers at centimetre wavelengths may provide one of the most rapid and reliable ways of identifying an episodic accretion event in high-mass star formation regions, and because of the maser monitoring, it has become a critical tool to investigate and study accretion bursts in star-forming regions \citep[e.g.][]{MacLeod:2018}. One such case of maser flaring has been observed toward a protostar in the star-forming region NGC6334I, which is situated in the ``mini-star burst'' region NGC6334 \citep{Wills:2013}. 

Early infrared continuum observations revealed that NGC6334 is one of the most active star-forming regions in the Milky Way \citep{Harvey:1983}. Subsequent observations with higher angular resolution have shown that this region houses many young, deeply embedded protostellar clusters \citep{Persi:2008, Feigelson:2009, Hunter:2006, Hunter:2014}. Its adjacent millimeter emission core, NGC6334I(N) has a submillimeter bolometric luminosity of $\sim$$1.7~\times~10^{4}$ L$_\odot$ \citep{Sandell:2000}. The distance to this object is determined to be $1.33\pm 0.12$ from its trigonometric parallax \citep{WuNGCParallax:2014, Chibueze:2014}. In this work, we likewise adopt that value as the distance to NGC6334I. Submillimetre observations by \citet{Hunter:2006} resolved the NGC6334I region into four compact millimetre continuum sources, corresponding to a previously well-known ultra-compact H{\sc ii} region (UCH{\sc ii}) which they label SMA3 (NGC6334F), two line-rich hot cores SMA1 and SMA2, and a dust core, SMA4. \citet{Brogan:2016} renamed them from SMA1..SMA4 to MM1..MM4. To ensure consistency with previous studies, we adopt the naming convention introduced by \citet{Brogan:2016} and refer to these millimetre continuum sources as `sub-regions' throughout this paper.

MM1 is situated at the origin of a high-velocity bipolar molecular outflow \citep{Leurini:2006, Beuther:2008, Qui:2011}. \citet{Brogan:2016} and  \citet{Hunter:2017}\ identify MM1 as the brightest in millimetre continuum emission and find that it contains multiple dust cores. On this basis, they further divided MM1 into 7 continuum sources, from MM1-A to MM1-G. They reported that among these 7, MM1-A, MM1-B and MM1-D are the most dominant, with MM1-B being the brightest. MM3 hosts the ultra-compact UCH{\sc ii} region NGC6334F (G351.42$+$0.64) \citep{Rodriguez:1982, Gaume:1987, Ellingsen:1996b}. 

Not long after the discovery of cosmic masers in 1966, 1665-MHz and 1667~MHz OH maser lines were reported in NGC6334 in 1968 \citep{Weaver:1968}. NGC6334I hosts a variety of maser sources; from hydroxyl (OH) \citep{Meeks:1969,MoranRodriguez:1980}, methanol (\methanol) at both 6.7-GHz \citep{Menten:1991} and 12-GHz \citep{Batrla:1987}, water (\water) at 22.2-GHz \citep{Meeks:1969} and a variety of other methanol transitions \citep{Haschick:1989, Haschick:1989b, Haschick:1990, Menten:1989, Ellingsen:2002}. Both OH and water maser emission have been detected towards MM3 \citep{Forster:1989, Hunter:2018, Chibueze:2021}, in addition to numerous class~II methanol maser transitions \citep{Cragg:2001}. MM3 has generally shown the strongest 6.7 class~II methanol maser activity in all previous studies \citep{Menten:1991, Ellingsen:1996b, Walsh:1998, Goedhart:2004, Hunter:2018} and houses one of the brightest 6.7-GHz methanol masers in the Milky Way. The strongest class~II methanol maser emission from MM3 is observed over the velocity range $-11.5$ -- $-9.5$ km~s$^{-1}$ and has been confirmed through interferometric imaging for the 6.7, 12.2, 19.9, 23.1, 37.7, 38.3 and 38.5-GHz transitions \citep{Norris:1993, Ellingsen:1996, Ellingsen:2018, Krishnan:2013}. Similarly, all subsequent observations since 1992 have reported strong class~II methanol maser detection associated with MM2 \citep{Norris:1993, Ellingsen:1996, Walsh:1998, Hunter:2018}.

Many of the maser transitions associated with NGC6334I have shown variability in their spectra since their initial detection \citep{MacLeod:2018, Hunter:2017, Hunter:2018, Brogan:2016, Brogan:2018}. The first interferometric images of the 6.7-GHz methanol maser emission in NGC6334I were made in 1992 by \citet{Norris:1993} and show two sites of emission separated by approximately 2 arcseconds. The observations in 1992 and 1993 reported slight variability in the 6.7-GHz methanol masers in MM3 (NGC6334F) \citep{Caswell:1995a}, but it was sufficiently unremarkable so that it was not included in the first paper investigating the variability of the 6.7-GHz transition \citep{Caswell:1995}. The Hartebeesthoek 26m radio telescope has been monitoring the 6.7-GHz methanol maser emission from NGC6334I every 10-14 days since February 1999 \citep{MacLeod:2018}. 

\citet{Goedhart:2004} reported that in February 1999, there was an increase in the flux density of the 6.7-GHz methanol maser emission at a velocity $-$5.88 km~s$^{-1}$, which peaked in Nov 1999. In 2015, \citet{MacLeod:2018} detected a flare in multiple maser species (methanol, hydroxyl, and water) associated with NGC6334I. The flare was most pronounced in the 6.7-GHz methanol masers, with some spectral components showing an increase in intensity by a factor of 20. The flare commenced in January 2015, peaking on 2015 August 15. Before the 2015 flaring event, there were no reported detections of 6.7-GHz methanol masers associated with MM1. Only MM2 and MM3 have hosted 6.7-GHz methanol masers before the 2015 flaring event. A recent study by \citet{Chibueze:2021} shows the results of multi-epoch VLBI observations of the 22-GHz H$_{2}$O maser emission using a combination of the VERA and Korean VLBI Network (KVN) to probe the kinematics of the gas surrounding MM1 and MM3 during November 2015 to January 2016. Their findings show that the H$_{2}$O maser proper motions in MM1 are mostly driven by the radio jet from MM1-B, and those in MM3 are largely driven by the expansion from MM3.

In this paper, we report milliarcsecond resolution observations of the 6.7-GHz methanol maser emission associated with NGC6334I at four epochs made with the Long Baseline Array (LBA). With these VLBI data, for the first time, we have information before, during and after a major methanol maser flare event at milliarcsecond resolution. These data, combined with the maser monitoring and multi-wavelength published data on NGC6334I, provide new information on the relationship between maser flares and episodic accretion events in high-mass star formation regions.

\section{Observations and Data Reduction}\label{sec:ObsandCal}

The LBA is a southern hemisphere VLBI array operated as a National Facility by the CSIRO and the University of Tasmania. There are typically five or six telescopes participating in most array experiments, with the exact number depending upon the source location, frequency of the observations and telescope availability. In this paper, we have included data from four LBA sessions spanning a period of 10 years, with two epochs in 2010 and one each in 2015 and 2020. The details of the observations are provided in Table~\ref{tab:Obsdetailsflaring}. In October 2015 and March 2020, the observations of other methanol maser sources were interleaved with those of NGC6334I, but we only report on the latter here. A total of eight different antennas participated in one or more of the sessions. The Australian Telescope Compact Array (ATCA)  was used in phased array mode in these observations, with the signals from five of the antennas combined prior to recording.

The data was recorded in dual circular polarisation with the $5_{1}-6_{0}$ A$^{+}$ transition of methanol (rest frequency 6.6685192 GHz) observed in the baseband data of one of the four 16 MHz IFs. The DiFX correlator was used to correlate the data \citep{Deller:2011}. A 2 MHz zoom band was correlated to cover the frequency range of the NGC6334I maser emission at each epoch with 2048 spectral channels, providing a channel spacing of 0.976 kHz (0.044 km~s$^{-1}$). The only exception was the session in March 2020, a 4 MHz zoom band with 4096 channels was utilised. The V255 experiments were primarily intended to make trigonometric parallax scans on southern 6.7-GHz methanol maser sources and so were scheduled such that observations of the maser target were interleaved with regular scans on nearby compact extragalactic continuum sources to provide phase referencing. For NGC6334I, the extragalactic continuum source selected for phase referencing proved unsuitable, which meant that relative astrometry between epochs could not be obtained and, consequently, the proper motions of the masers could not be investigated.  \citet{Chibueze:2014} presented successful phase$-$referencing observations at 22-GHz for H$_{2}$O masers in NGC6334I(N), using the compact continuum source J1713$-$3418 as the phase reference. At 6.7-GHz, J1713$-$3418 had a peak intensity of $\sim$0.1 Jy$~$beam$^{-1}$ \citep{Petrov:2025}. This could potentially be used to do absolute astrometry for future 6.7~GHz observations. Table~\ref{tab:Obsdetailsflaring} lists the details of the 6.7-GHz methanol maser observations between March 2010 and March 2020, including the start time and the duration of the observations.

\smallskip


%

\smallskip

The correlated data of all four epochs were calibrated using the Astronomical Image Processing System (AIPS) software using the common spectral-line VLBI data reduction technique. Flag files from each antenna were applied to exclude data from the time ranges when they were not on the source or when there were other issues with the data quality noted during the observations. Corrections for ionospheric delays, Earth Orientation Parameters and the changing parallactic angle over the course of the observations were applied by using the AIPS ionospheric calibration (VLBATECR) and CLCOR tasks. The Doppler velocity corrections were applied to the data using the AIPS task CVEL, to reposition the maser spectrum in the bandpass and to correct for variation in the Doppler shifts due to the Earth's rotation and orbit. We utilised autocorrelation data of the 6.7-GHz methanol masers observed with the Hartebeeshoek radio telescope to calibrate the amplitude for each epoch of the observations (assuming 3830 Jy for the March 2010 and July 2010, 4375 Jy for the October 2015 and 4094 Jy for the March 2020). Cross-correlation amplitude correction, using the AIPS task ACCOR, was applied to correct the data amplitude for incorrect sampler statistics and system equivalent flux density (SEFD). The average uncertainties in the intensity scales for the four epochs were 0.32, 0.29, 0.56 and 0.38 Jy$~$beam$^{-1}$ respectively. The 'template spectrum method' was used to calibrate relative antenna gains using a reference autocorrelation spectrum, correcting for residual amplitude differences between VLBI antennas. The uv-data splitting task SPLIT was used to extract a single autocorrelation spectrum scan with ATCA (the most sensitive telescope in the array), and then task ACFIT was used to determine antenna gains of all antennas and for all time ranges relative to this scan. The resultant amplitude corrections or gains were then applied to the cross-correlation maser data for each baseline. Fringe fitting was performed using the AIPS task FRING on a strong, compact calibrator source to determine and correct for instrumental/electronic delays and phase offsets, and the solution was applied to the maser dataset. FRING was then used on the same maser spot (spectral velocity channel) to determine the time-dependent fringe rates. The local-standard-of-rest (LSR) velocity of the maser spot used for the fringe fitting in all four epochs was 9.65 km~s$^{-1}$. This compact maser emission was then self-calibrated iteratively using the AIPS task CALIB, and the solutions were applied to the whole maser data. Finally, spectral cubes with stokes $I$ were made using the AIPS task IMAGR, with a channel spacing of 0.976 kHz (0.044 km~s$^{-1}$).

Table~\ref{tab:Obsdetailsflaring} lists the details of the observations and other critical information such as project code, total maser emission region, polarisation, bandwidth, spectral resolution, and synthesised beam size.

\smallskip

\begin{table*}
\centering
\caption{Details of the LBA observations and the information about the obtained data of 6.7-GHz methanol maser sources in NGC6334I.. Column 1 lists the different parameters of the analysed image data cube; columns 2-5 list the corresponding values of those parameters for epochs 1-4. The participating stations are the Australia Telescope Compact Array~(AT), Ceduna~(CD), Katherine~(KE), Hartebeesthoek~(HH), Hobart~(HO), Mopra~(MP), Parkes~(PA) and Warkworth~(WA).}\label{tab:Obsdetailsflaring}
\begin{tabular}{rccccl}\toprule \toprule
\multicolumn{1}{c}{\bf Parameters} &
\multicolumn{1}{c}{\bf Epoch 1}  &
\multicolumn{1}{c}{\bf Epoch 2}  &
\multicolumn{1}{c}{\bf Epoch 3}  &
\multicolumn{1}{c}{\bf Epoch 4}  \\
\hline
\hline
Project Code                           & V255I                       & V255J                      & V255Z                      & V255AI                 \\
Observing Date                         & 09 March 2010               & 22 July 2010               & 01 October 2015            & 19 March 2020          \\
Modified Julian Date                   & 55264                       & 55399                      & 57296                      & 58927                  \\
Observation Start Time (UTC)            &14:13                        & 08:11                      & 22:00                      & 05:26                  \\
On-Source Time (Hrs)                          & 1.4                          & 1.7                         & 0.8                         & 1.0                     \\
Participating Antennas                 & AT, CD, HO, HO, PA          & AT, CD, HO, MP, PA         & AT, CD, HH, HO, MP, WA     & AT, CD, KE, MP, PA, WA   \\
Maser Emission Regions                 & MM2, MM3                    & MM2, MM3                   & MM1(I,II,III), MM2, MM3       & MM1(I,II), MM2, MM3   \\
Polarization products                  & dual circular               & dual circular              & dual circular              & dual circular            \\
Bandwidth                              & 2-MHz                       & 2-MHz                      & 2-MHz                      & 4-MHz                    \\
Imaged Field Width (arc sec)           & 4.096 x 4.096               & 4.096 x 4.096              & 4.096 x 4.096              & 4.096 x 4.096            \\
Cube Spectral Resolution (km~s$^{-1}$)  & 0.044 km~s$^{-1}$             &  0.044 km~s$^{-1}$           &  0.044 km~s$^{-1}$           & 0.044 km~s$^{-1}$          \\
Synthesized Beam (" mas $\times$~mas " (P.A.$^{o}$ )) & 5.43 x 4.45 (68.50$^{o}$)  & 4.99 x 4.76 (44.32$^{o}$) & 3.82 x 3.05 (-63.18$^{o}$) & 5.00 x 3.54 (19.44$^{o}$) \\
Number of channels in spectral data    & 2048                        &   2048                     &    2048                    & 4096                     \\
RMS noise per channel (Jy$~$beam$^{-1}$) &  0.357, 1.83, 0.043         &  0.322, 1.38, 0.043        &   0.566, 2.9, 0.118        &  0.308, 1.14, 0.102      \\
(median, max, min)                     &                             &                            &                            &                          \\
\bottomrule
\hline
\end{tabular}%
\end{table*}

\section{Results}\label{FlaringResults}

This paper presents NGC6334I 6.7-GHz methanol maser images at milliarcsecond resolution at four epochs, which span a time range between March 2010 and March 2020. Of these, the October 2015 observations were conducted when the NGC6334I region was experiencing a flaring event. 

\begin{figure}
\centering
\includegraphics[scale=0.45,width=0.5\textwidth,height=\textheight,keepaspectratio]{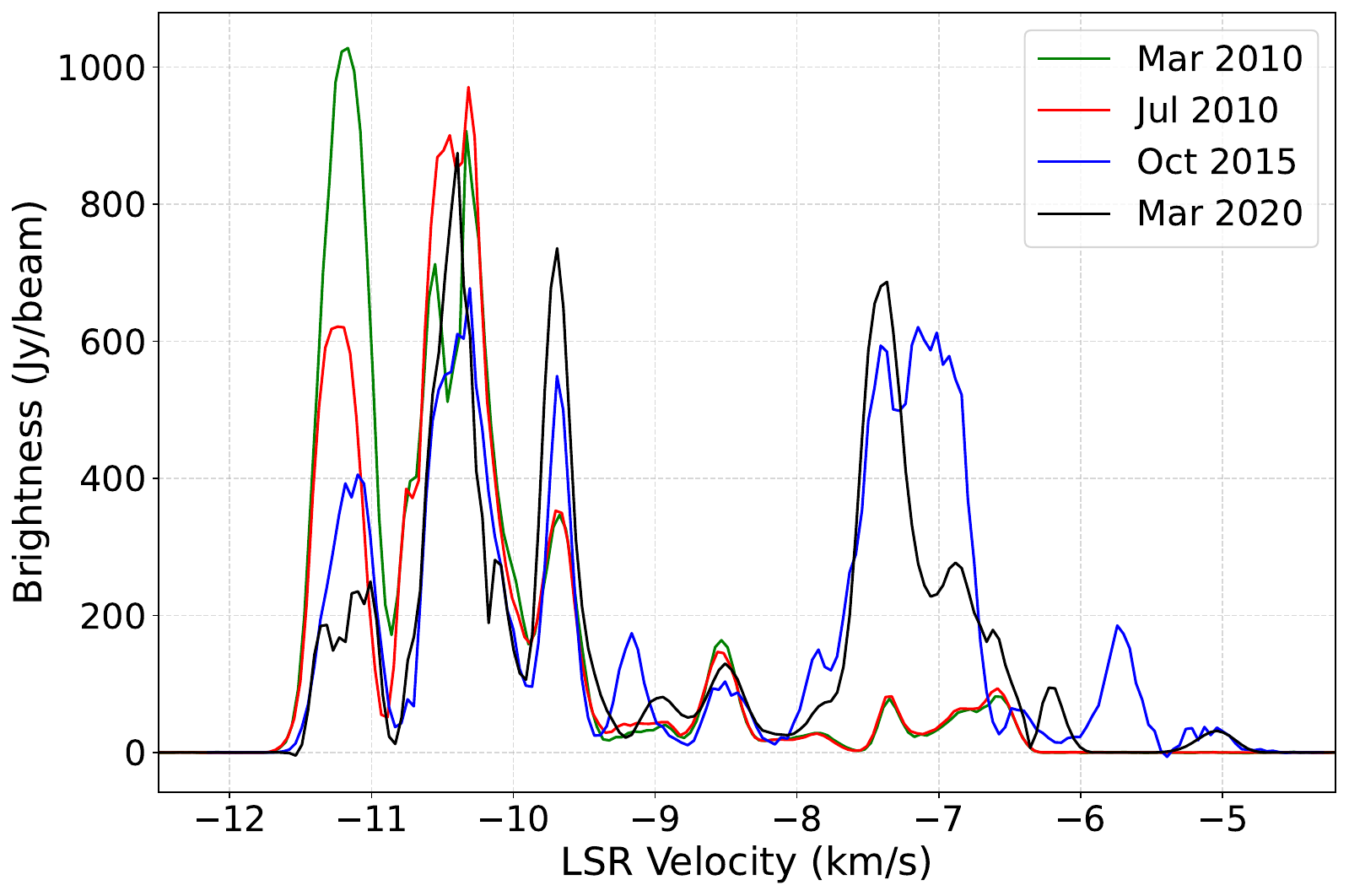}
\caption{Interferometric total-power (Stokes-I) spectra, representing the integrated maser emission over the full NGC6334I region, extracted from the image cubes for all four epochs. Each epoch's emission spectrum is colour-coded and labelled in the figure.}
\label{fig:EmissionSpectraAllEpochs}

\end{figure}

\begin{figure}
\centering
\includegraphics[scale=0.45,width=0.5\textwidth,height=\textheight,keepaspectratio]{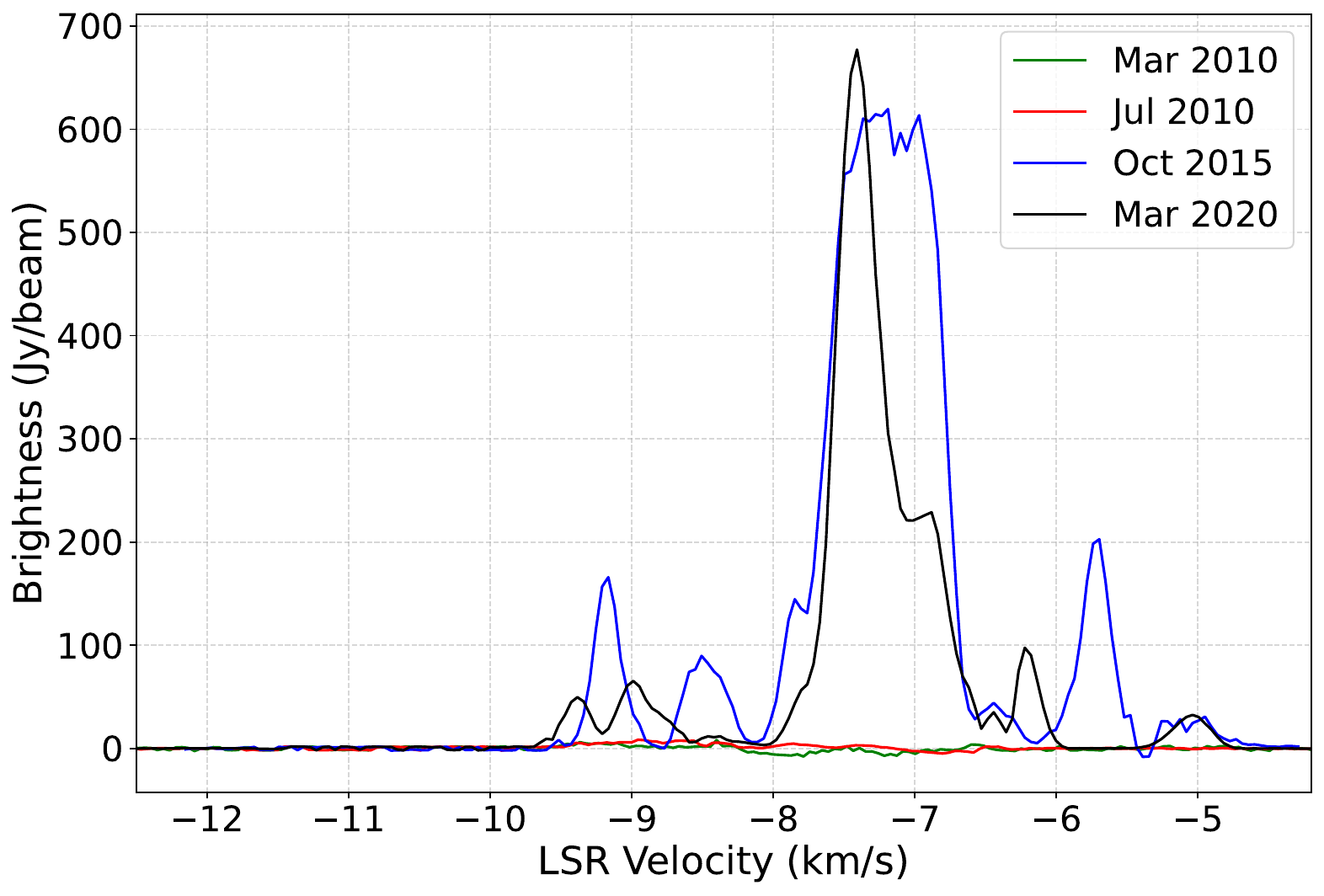}
\caption{Same as Figure~\ref{fig:EmissionSpectraAllEpochs}, but from sub-region MM1}
\label{fig:brightnessspectraMM1}
\end{figure}

\begin{figure}
\centering
\includegraphics[scale=0.45,width=0.5\textwidth,height=\textheight,keepaspectratio]{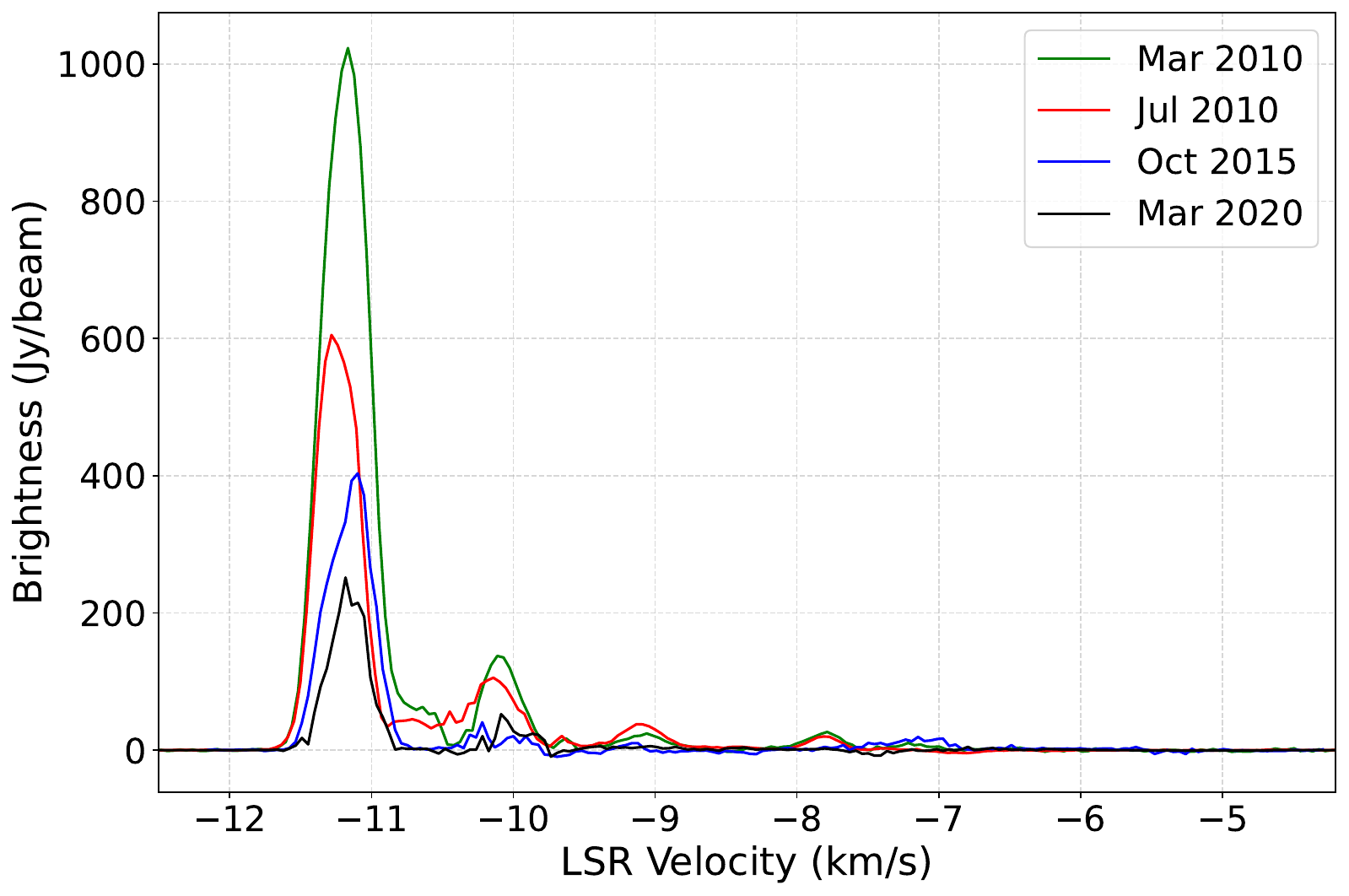}
\caption{Same as Figure~\ref{fig:EmissionSpectraAllEpochs}, but from sub-region MM2}
\label{fig:brightnessspectraMM2}
\end{figure}

\begin{figure}
\centering
\includegraphics[scale=0.45,width=0.5\textwidth,height=\textheight,keepaspectratio]{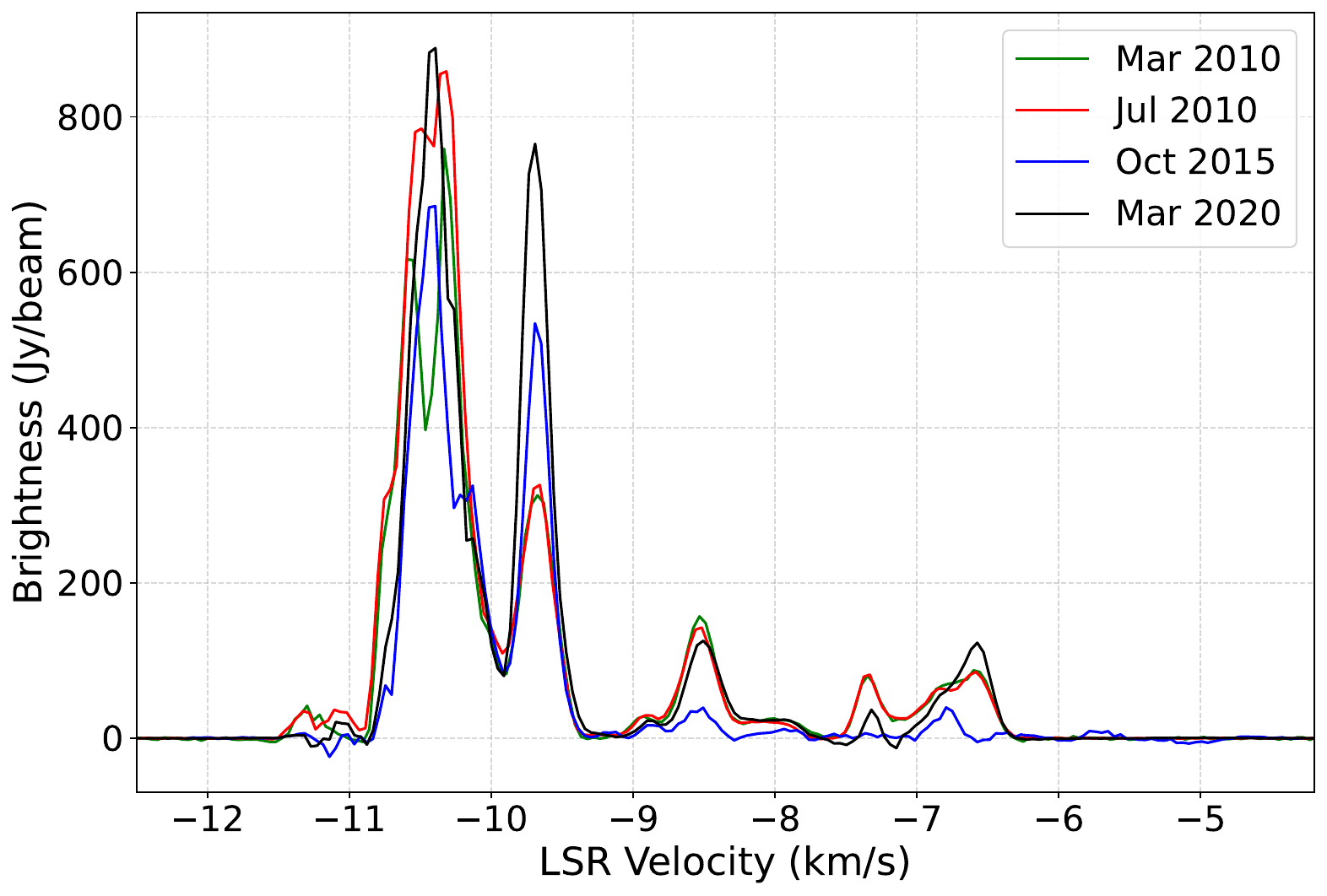}
\caption{Same as Figure~\ref{fig:EmissionSpectraAllEpochs}, but from sub-region MM3}
\label{fig:brightnessspectraMM3}
\end{figure}

Figure~\ref{fig:EmissionSpectraAllEpochs} shows a comparison of the 6.7-GHz methanol maser emission spectra for the whole NGC6334I region extracted from the image cubes for all four epochs. It shows that for the 2010 epochs, the strongest emission was in the velocity range $-10$ to $-11.5$ km~s$^{-1}$ and was around an order of magnitude stronger than that observed in the velocity range $-8.0$ to $-6.5$ km~s$^{-1}$. The image-cube spectra in 2015 and 2020 show significant differences at the most negative velocities with declined emission and the velocity range of $-8.0$ to $-6.5$ km~s$^{-1}$ with emission comparable in intensity to that in the velocity range of $-10$ to $-11.5$ km~s$^{-1}$. This is consistent with the single dish monitoring results of \citet{MacLeod:2018} which show persistent changes in the 6.7-GHz methanol maser spectrum following the approximately year-long flare event, resulting in maser detections within an area of 5$\times$7 arcseconds. Figures~\ref{fig:brightnessspectraMM1}-\ref{fig:brightnessspectraMM3} show a comparison of the interferometric maser brightness spectra for all three NGC6334I sub-regions: MM1, MM2 and MM3 for all four epochs.

\begin{figure}
\centering
\includegraphics[scale=0.2,width=0.5\textwidth,height=\textheight,keepaspectratio]{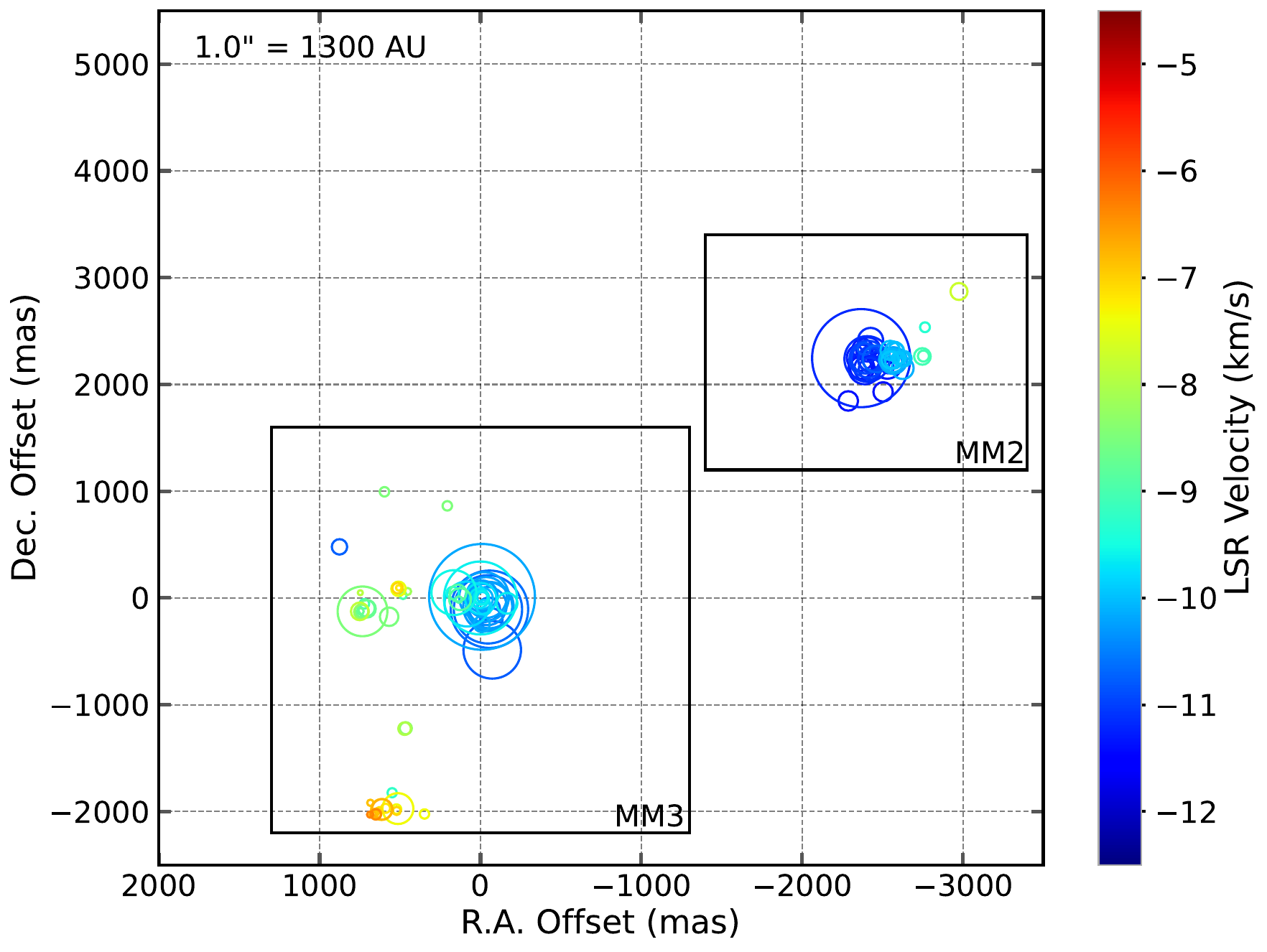}
\caption{Interferometric spot maps of 6.7-GHz methanol maser features in NGC6334I in March 2010. All maser features are plotted on the Right Ascension (R.A.) and Declination (DEC.) map with their velocity information. The whole region is further divided into multiple sub-regions based on the nomenclature of \citet{Brogan:2016, Hunter:2017, Hunter:2018}. Note that there was no emission detected from the MM1 during this epoch, which can also be seen from the lack of any spectral components in Figure~\ref{fig:brightnessspectraMM1}.}
\label{fig:maserfeaturesI}
\end{figure}

\begin{figure}
\centering
\includegraphics[scale=0.2,width=0.5\textwidth,height=\textheight,keepaspectratio]{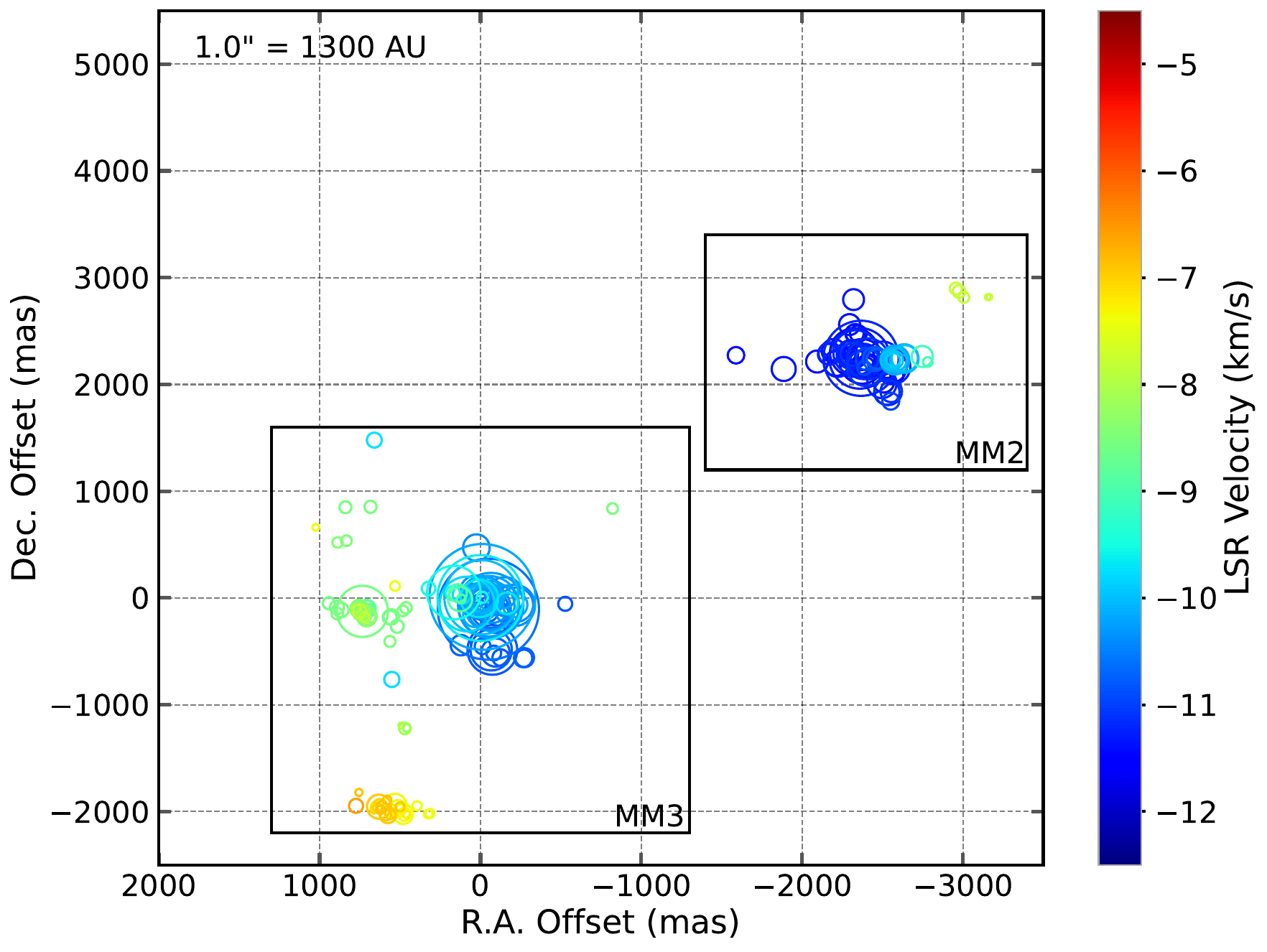}
\caption{Same as Figure~\ref{fig:maserfeaturesI} but in July 2010. Similar to March 2010, there was no emission detected from the MM1 for this epoch as well.}
\label{fig:maserfeaturesJ}
\end{figure}

\begin{figure*}
\centering
\includegraphics[scale=0.9,width=\textwidth,height=0.65\textheight, keepaspectratio]{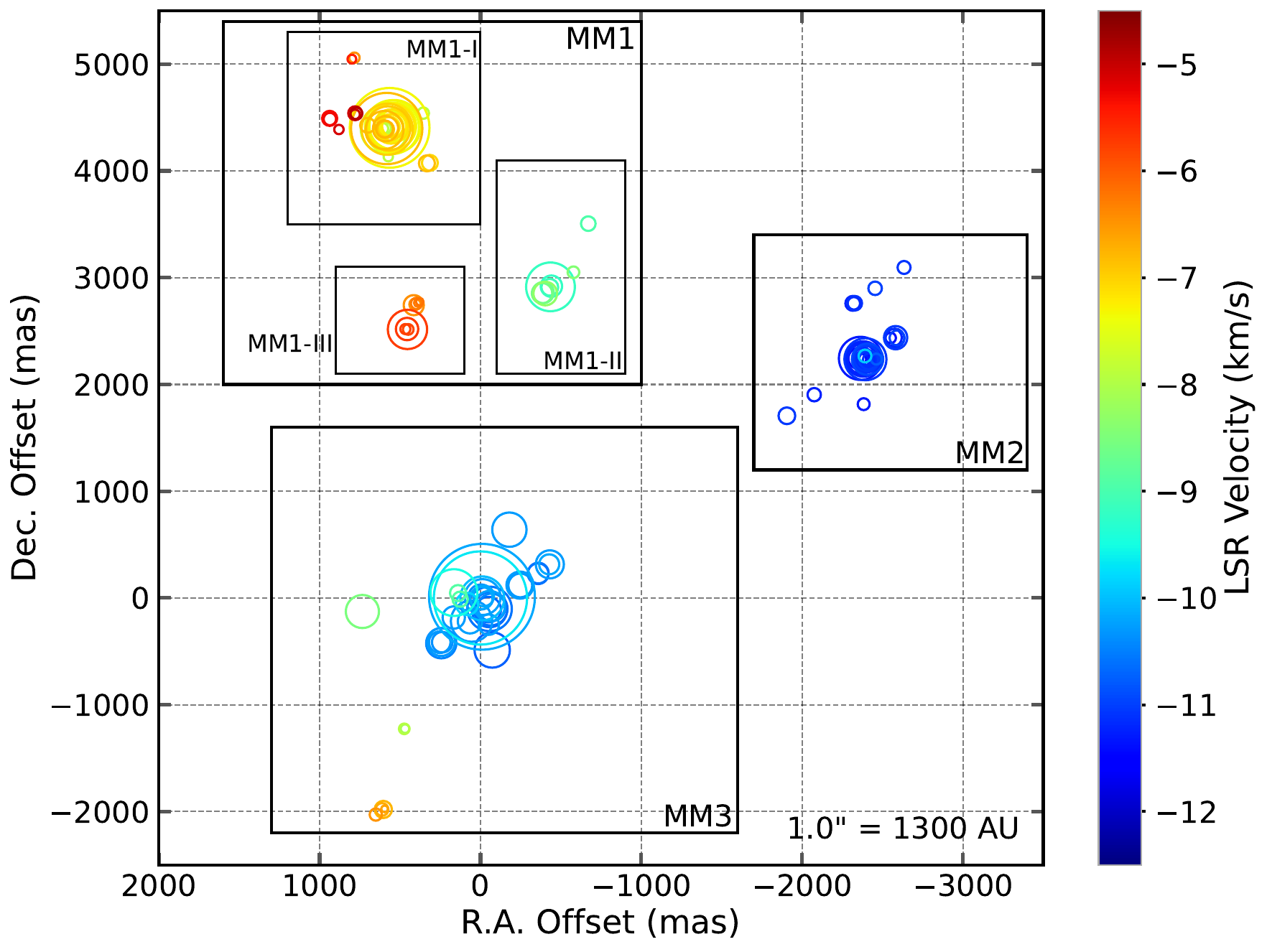}
\caption{Same as Figure~\ref{fig:maserfeaturesI} but in October 2015.}
\label{fig:maserfeaturesZ}
\end{figure*}

\begin{figure*}
\centering
\includegraphics[scale=0.95,width=\textwidth,height=0.6\textheight, keepaspectratio]{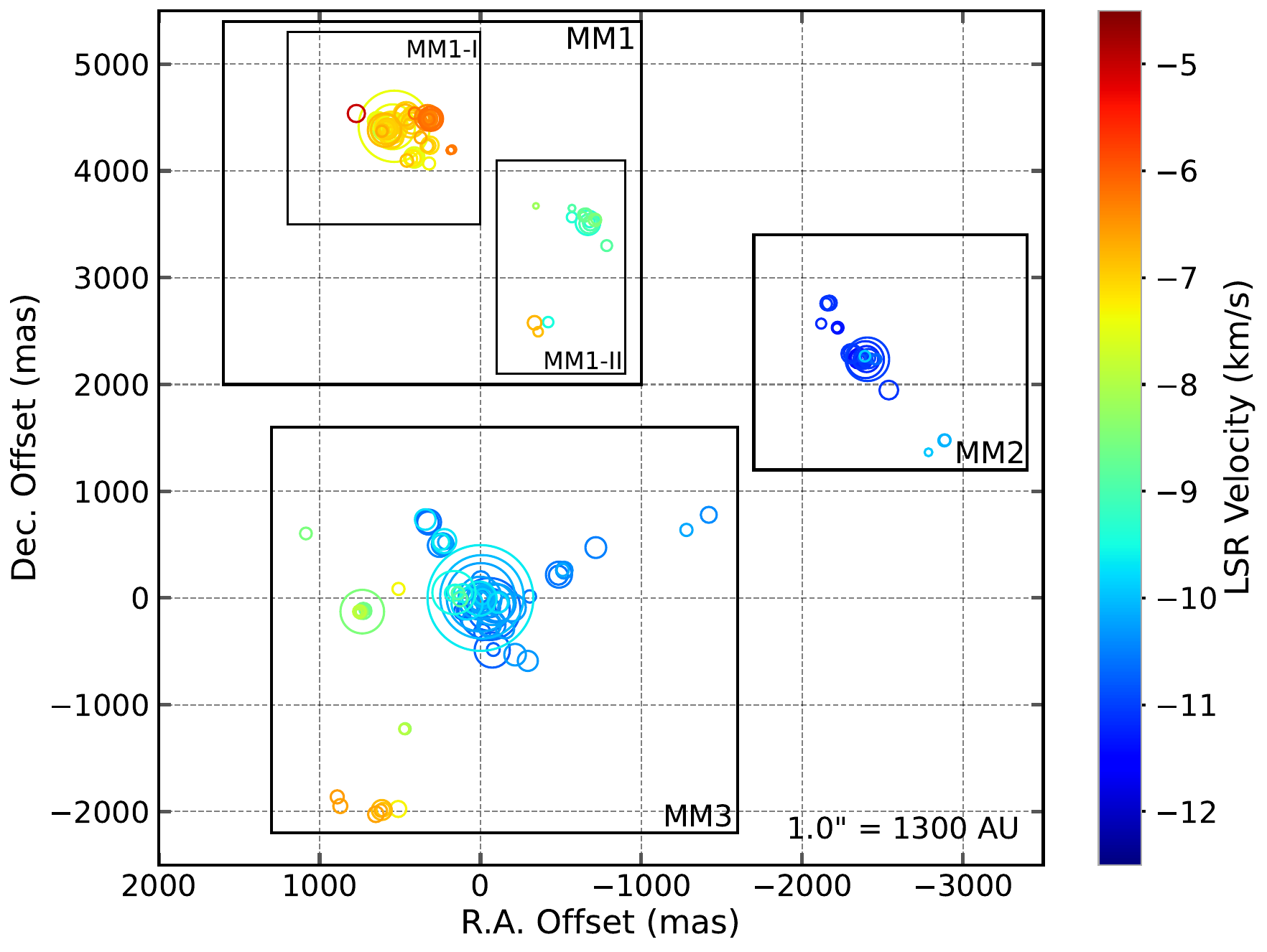}
\caption{Same as Figure~\ref{fig:maserfeaturesI} but in March 2020.}
\label{fig:maserfeaturesAI}
\end{figure*}

We have imaged a region of NGC6334I covering 16$\times$16 arcsecond, from a Right Ascension (J2000) of 17:20:53.7 to 17:20:53 and from a Declination $-$35:46:54.0 to $-$35:47:04.0. Figures~\ref{fig:maserfeaturesI}--\ref{fig:maserfeaturesAI} show the VLBI interferometric spot maps of the 6.7-GHz methanol masers for each of the four epochs, from March 2010 to March 2020. The large angular extent of the 6.7-GHz methanol masers in NGC6334I presents some challenges for imaging at milliarcsecond resolutions. Initial images with a large cell-size were created to identify the total extent and locations of the maser emission clusters, and this was followed by simultaneous imaging in CLEAN deconvolution with an optimal cell-size of up to 3 fields, which covered all the emission apparent in the initial images. This approach is critical to accurately image masers for sources where there is strong emission from different regions which overlap in velocity. In some cases, the multi-field higher resolution images revealed weaker maser emission towards the edge of a field. When this occurred, the field centre or extent was modified in an iterative manner until we were confident that all the maser emission was captured. 

The combined effect of different sensitivities of antennas in the array, on-source time, and maser intensity implies that the noise in the image cubes changes with epoch and velocity. We are interested in the changes in the location and intensity of the maser emission during and after the flare compared to those before the flare. To facilitate that comparison, we have used the same objective criteria to create the spot maps for each epoch. The criteria are that the emission has a signal-to-noise ratio (SNR) greater than 8; an intensity greater than 1 Jy$~$beam$^{-1}$ and be present in 3 consecutive velocity channels. The spot maps were created by fitting a 2D Gaussian brightness component to all components in each channel of the image cubes with a peak brightness $>8\sigma$. Each circle in the spot map represents a maser feature and is centred at the weighted average of the brightness peaks of the 2D Gaussian components located within 0.2 mas and 0.5 km~s$^{-1}$ spatial and velocity range, respectively. The diameter of the circle is proportional to the integrated flux density. The colour of the circles depends on the velocity of the maser feature. Tables~\ref{tab:Detailsv255Iallmaserfeatures}--~\ref{tab:Detailsv255AIallmaserfeatures} in Appendix~\ref{appendixA:ngc6334bigtables} lists all the identified maser features for the whole NGC6334I region (MM1, MM2, MM3) for each epoch. For each maser feature, the right ascension and declination offsets from the centre position (R.A. = 17:20:53.454, DEC. = $-$35:47:00.521 [J2000]), velocity, integrated flux density and the sub-region (MM1, MM2 or MM3) it is associated with, are specified in the table.

\smallskip

\smallskip

\smallskip

\smallskip

Figures~\ref{fig:maserfeaturesI}--\ref{fig:maserfeaturesAI} show 6.7-GHz methanol maser emission that is associated with each of the three-millimetre continuum sources (MM1, MM2 and MM3) in NGC6334I at the four epochs. There is emission associated with MM2 and MM3 across all four epochs, whereas MM1 has maser emission in only the two epochs taken after the commencement of the 2015 flare. In the following sections, we discuss the emissions associated with each of these three sub-regions in detail.

\subsection{Sub-Region MM3}\label{subregionMM3}

MM3 is associated with the well-studied ultra-compact H{\sc ii} region NGC6334F \citep{Rodriguez:1982,Ellingsen:1996,Hunter:2006,Brogan:2016}. Figure~\ref{fig:brightnessspectraMM3} shows that for all four epochs, there is strong 6.7-GHz methanol maser emission from MM3 in the velocity range from $-$11 km~s$^{-1}$ to $-$9.3 km~s$^{-1}$. After the commencement of the flare (October 2015 and March 2020), no emergence of any new spectral components was found from MM3, although the relative intensity of the emission has changed significantly over the 10-year duration of these observations. The emission at the most negative velocities shows a significant decrease in intensity following the onset of the flare, while the maser components with peak velocities of approximately $-$10.4 and $-$9.7  km~s$^{-1}$ increase in intensity. The latter of these is now the brightest milliarcsecond-scale emission in the 6.7-GHz transition associated with NGC6334I, and in October 2020 had an intensity approximately a factor of four stronger than that found a decade earlier. The emission around $-$8.6 km~s$^{-1}$ shows relatively little variation in intensity over the four epochs; therefore, this emission was used for the self-calibration at each epoch.

In the interferometric maser spot maps shown in Figures~\ref{fig:maserfeaturesI}--\ref{fig:maserfeaturesAI}, MM3 hosts the 6.7-GHz class~II methanol emission covering an area on the sky of approximately 2.3$~\times~$2.3 arcseconds. The strongest emission in the velocity range from $-$11 km~s$^{-1}$ to $-$9.3 km~s$^{-1}$ is located towards the western edge of the cometary ultra-compact H{\sc ii} region. Emission at velocities around $-$8 km~s$^{-1}$ is located approximately 0.75 arcseconds (1000 au) eastwards, and the most positive velocity emission in the pre-flare ($-$7.5 km~s$^{-1}$ to $-$6.5 km~s$^{-1}$) is located 2 arcseconds (2600 au) south. A comparison of the 6.7-GHz methanol masers associated with MM3 over the four epochs shows that both the number of components and their angular extent were at their lowest in October 2015, at the peak of the flare. This may be partly attributed to the higher noise level in the October 2015 data, nearly a factor of two greater than in the other three epochs, making it more likely that weaker maser emission was missed. Nevertheless, approximately 90\% of the detected emission in all epochs is above the sensitivity limit in October 2015. Therefore, variations in the noise levels are unlikely to have significantly affected the number of detected components. However, the data are consistent with the maser emission in MM3 being partly suppressed or reduced at the same time when it peaked in MM1. The strongest 6.7-GHz methanol maser emission in MM3 has an offset of approximately 3 arcseconds from the peak of the MM1 continuum emission, corresponding to a light-travel time of around 20 days at the assumed distance of 1.3~kpc. This suggests that changes in the radiation field of MM1 due to the 2015 flare may be impacting the masers in MM3 during October 2015. The single dish monitoring data \citep[see figure 1 of][]{MacLeod:2018} does not show evidence of significant time delays across the spectrum. This suggests that any changes which impact the three 6.7-GHz methanol maser emission regions in NGC6334I either happened at nearly the same time, too fast to be resolved by the approximately fortnightly observation cadence of \citeauthor{MacLeod:2018}, or were independent.

\subsection{Sub-Region MM2}\label{subregionMM2}

MM2 was first identified as a distinct site of maser emission with the detection of 12.2-GHz methanol masers at this location by \citet{Norris:1988}. Figure~\ref{fig:brightnessspectraMM2} shows that for all four epochs, there is strong emission in the velocity range from $-$11.6 km~s$^{-1}$ to $-$10.8 km~s$^{-1}$. This emission was strongest in March 2010 and has shown a consistent decline in intensity in subsequent epochs, although the largest decline occurred well before the 2015 flare event. Along with the strong emission, there is a weaker emission present in the velocity range from $-$10.4 km~s$^{-1}$ to $-$9.7 km~s$^{-1}$, most of which has significantly declined in October 2015 but appears to have slightly strengthened in March 2020. There is also very weak emission in velocity ranges from $-$9.3 km~s$^{-1}$--$-$8.8 km~s$^{-1}$ and from $-$8.0 km~s$^{-1}$--$-$7.6 km~s$^{-1}$ in both 2010 epochs, which has disappeared in October 2015 and March 2020.

As shown in Figures~\ref{fig:maserfeaturesI}--\ref{fig:maserfeaturesAI}, MM2 hosts 6.7-GHz class~II methanol emission over a 1.5$~\times~$2.5 arcsecond region, which is located approximately 3.25 arcseconds (4200 au) north-west of the strongest methanol maser emission in MM3. Figures~\ref{fig:maserfeaturesI}--\ref{fig:maserfeaturesAI} show significant differences in the distribution of the weaker maser emission in MM2 from epoch to epoch. The strongest 6.7-GHz methanol masers in MM2 and MM3 overlap in velocity, and some of the changes in maser distribution are likely related to this (even though the interferometric resolution of our observations is several orders of magnitude higher than the separation between the two sub-regions). This is particularly evident during the 2015 outburst event. The March 2020 maser distribution for MM2 also shows a significant difference from that seen in pre-flare epochs; however, we cannot determine from our data the degree to which the changes in MM2 are related to external influences (i.e. MM1) or changes in MM2 itself.

\subsection{Sub-Region MM1}\label{subregionMM1}

Prior to the 2015 flaring event, there had been no 6.7-GHz class~II methanol maser emission detected toward MM1 \citep{Ellingsen:1996}. We have imaged MM1 for all four epochs and Figure~\ref{fig:brightnessspectraMM1} confirms that in 2010 there was no associated 6.7-GHz methanol maser emission on milliarcsecond scales with an intensity greater than 1 Jy$~$beam$^{-1}$ in the LSR velocity range from $-$12.0 km~s$^{-1}$ to $-$4.5 km~s$^{-1}$. In contrast, in October 2015, 6.7-GHz methanol maser emission associated with MM1 was observed over a wide velocity range. The emission is strongest in the velocity ranges $-$8 km~s$^{-1}$ to $-$6.5 km~s$^{-1}$, with the highest intensity emission peaking at a velocity of around $-$7.5 km~s$^{-1}$. There are five spectral components in the velocity range $-$9.5 km~s$^{-1}$ to $-$8.9 km~s$^{-1}$, $-$8.8 km~s$^{-1}$ to $-$8.1 km~s$^{-1}$, $-$8.0 km~s$^{-1}$ to $-$6.5 km~s$^{-1}$, $-$6 km~s$^{-1}$ to $-$5.5 km~s$^{-1}$ and $-$5.4 km~s$^{-1}$ to $-$4.8 km~s$^{-1}$. The single dish monitoring observations of \citet{MacLeod:2018} found that commencing in January 2015, methanol emission in this velocity range in the NGC6334I region started flaring. Our LBA observation in October 2015 was conducted at approximately the time when the 6.7-GHz methanol maser flare was at its peak intensity.

As shown in Figures~\ref{fig:maserfeaturesI}--\ref{fig:maserfeaturesAI}, MM1 shows 6.7-GHz methanol emission covering a 3$~\times~$3-arcsecond region, north of MM3. \citet{Brogan:2016} reported that this sub-region houses the brightest millimetre continuum emission in the protostellar cluster. \citet{Hunter:2018} made VLA A-array observations of the 6.7-GHz methanol masers in NGC6334I in October and November 2016 (approximately a year following our LBA observations). Their observations show the strongest 6.7-GHz methanol maser emission associated with MM1 is located at the northern edge of the millimetre continuum emission, with the other main maser clusters to the west in the general direction of MM2.

In our October 2015 observations, we identified three main maser clusters of 6.7-GHz methanol masers in MM1, which we labelled MM1-I to MM1-III. Figures~\ref{fig:maserfeaturesZ}--\ref{fig:maserfeaturesAI} show these three maser clusters in rectangular boxes. The centre of the MM1-I hosts the strongest emission from this sub-region. It has offsets of $\sim~$580 mas and 4400 mas in right ascension and declination, respectively, from the centre of the MM3.

While the flare in the 6.7-GHz methanol masers associated with MM1 had its peak intensity around August 2015, the intensity of emission for part of its velocity emission range has subsequently declined. On the other hand, Figure 1 of \citet{MacLeod:2018} shows post-flare emission in the $-7.5$ -- $-6.5$ km s$^{-1}$ remained substantially stronger than the pre-flare levels through to late 2017 (the end of the observations reported in that paper). Figure~\ref{fig:maserfeaturesAI} shows that much of the 6.7-GHz methanol maser emission detected near the peak of the flare in 2015 is still detected five years later in March 2020. At that time, while the emission of the main feature has slightly became stronger, the emission in $-$9.5 km~s$^{-1}$ -- $-$8.8 km~s$^{-1}$ has significantly reduced and emission in the velocity range $-$8.8 km~s$^{-1}$ -- $-$8.1 km~s$^{-1}$ and $-$6.0 km~s$^{-1}$ -- $-$5.5 km~s$^{-1}$ has died down. 

Interestingly, the LBA observation in March 2020 found some new emission at around $-$6.2 km~s$^{-1}$. The main differences are in the intensity and location of the components south and west of the emission in MM1. MM1-I shows maser emission over the velocity range $-$7.5 km~s$^{-1}$ to $-$5.5 km~s$^{-1}$. Most of the maser emission in MM1 persisted in March 2020 (Figure~\ref{fig:maserfeaturesAI}). The maser emission in maser clusters MM1-II and MM1-III shows the most variability between October 2015 and March 2020. The MM1-II region has 6.7-GHz methanol maser emission over the velocity range $-$9.5 km~s$^{-1}$ -- $-$9 km~s$^{-1}$ and is situated southwest of MM1-I. The maser emission associated with the MM1-II in March 2020 shows significant evolution from that observed in October 2015. In March 2020, the northern maser cluster in MM1-II was active over an extended velocity range. There was maser emission in the middle of MM1-III at velocities around $-6$ km s$^{-1}$, but it disappeared in March 2020. On the other hand, maser emission in the southern region of MM1-II increased significantly at velocities around $-9$ km s$^{-1}$ in March 2020. A new maser emission also appeared in the northern region of MM1-II at velocities around $-6.5$ km s$^{-1}$. Between October 2015 and March 2020 epochs, MM1 showed far more substantial changes than the other regions.

The spectral channels in the velocity range $-$11.5 km~s$^{-1}$ to $-$10.0 km~s$^{-1}$ are affected by sidelobes arising from strong emission in the neighbouring MM2 and MM3. Due to the large field size and the complexity of the emission structure, the CLEAN deconvolution process had inherent limitations and was unable to fully mitigate sidelobe contamination at these velocities in MM1. However, even below such contamination levels, we find no evidence of 6.7-GHz methanol maser emission in MM1 in this velocity range before 2015.

\subsection{Comparison with Earlier Studies}\label{hunterminecomparion}

\begin{figure*}
\centering
\includegraphics[scale=0.95,width=\textwidth,height=0.6\textheight, keepaspectratio]{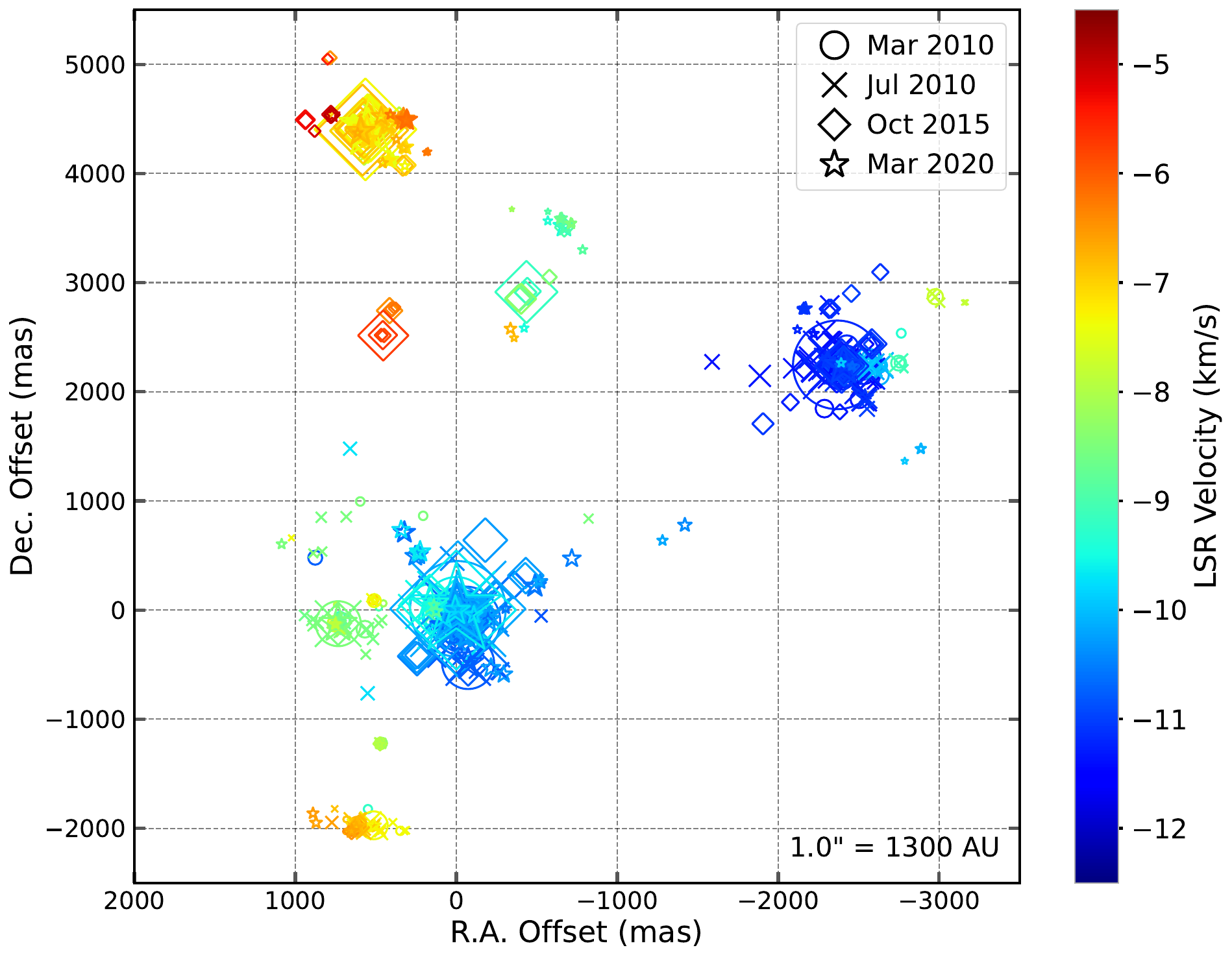}
\caption{Interferometric map of the total maser emission in NGC6334I star-forming region for all four epochs. Data for each epoch is plotted with a different marker, which is shown in the top left corner of the map.}
\label{fig:maserfeaturesAllepochs}
\end{figure*}

\begin{figure*}
\centering
\includegraphics[scale=0.2,width=\textwidth,height=\textheight,keepaspectratio]{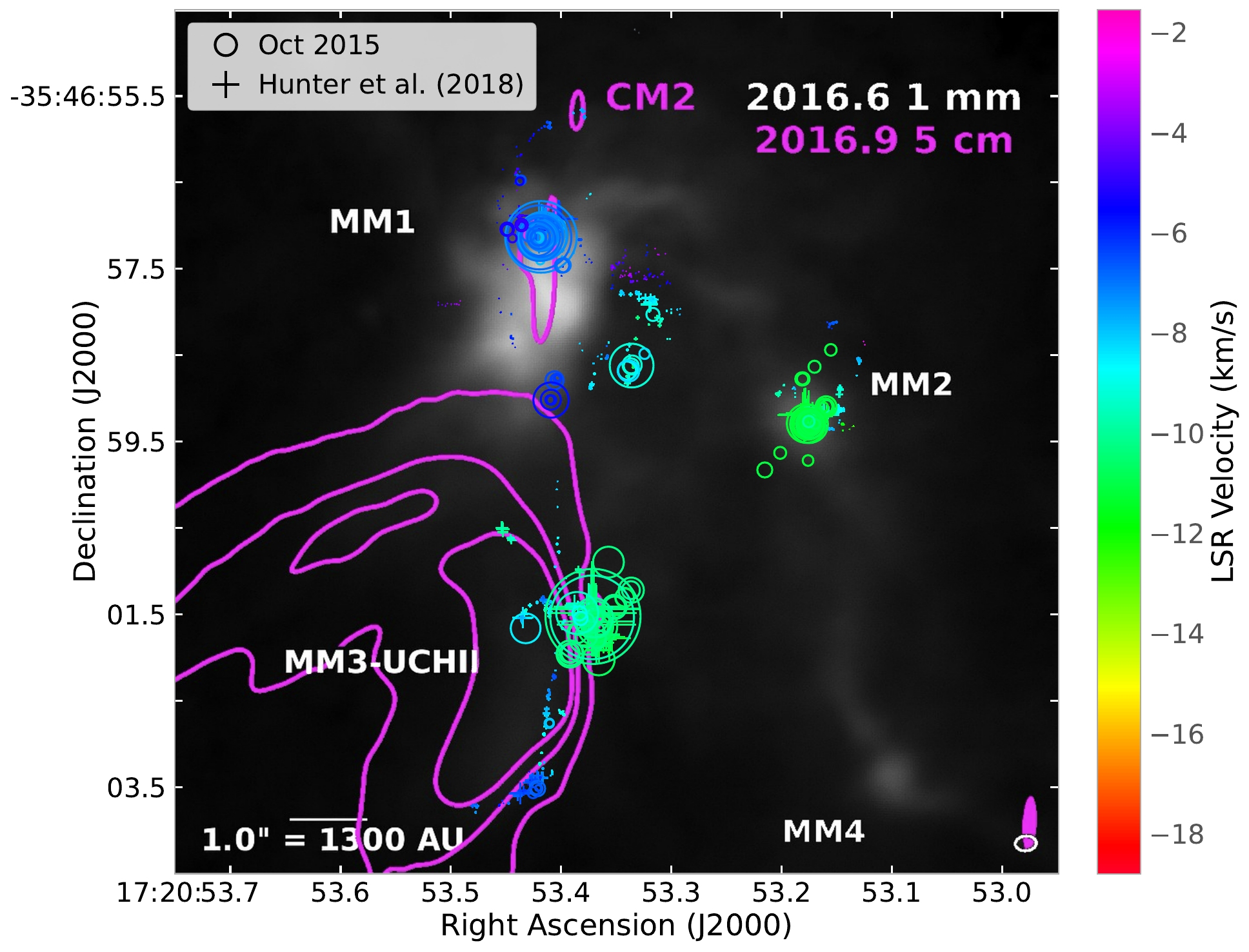}
\caption{Maser features in the total maser emission in NGC6334I star-forming region in October 2015, and maser features reported by \citet{Hunter:2018} at 6.7-GHz are overlaid on the ALMA 1 mm continuum image in 2016.6 in grayscale \citep{Hunter:2017}. The diameter of the symbol indicates the maser intensity, while the colour of the symbol indicates the LSR velocity. To facilitate better comparison, the distribution of the maser features in October 2015 has been shifted to align with the VLA map. Magenta contours show the 5 cm continuum emission in 2011.4 \citep{Brogan:2016} at levels of (4, 260, and 600) $\times$ the rms level of 3.7$~\times~$10$^{-2}$ mJy beam$^{-1}$. The millimetre continuum sources are labelled for reference. Map registration was done by eye; a shift of $\sim$10 mas produces noticeably poor alignment between the maps.}
\label{fig:maserfeaturesV255ZandHunter2018}
\end{figure*}

Figure~\ref{fig:maserfeaturesAllepochs} shows the 6.7-GHz methanol maser emission in NGC6334I across all four epochs and shows how it has changed. In particular, it highlights the significant changes in the maser clusters to the south and south-west of MM1 between 2015 and 2020 and the change in the extent of the MM2 and MM3 emissions over the 10-year period. Figure~\ref{fig:maserfeaturesV255ZandHunter2018} compares 6.7-GHz methanol maser emission from the LBA October 2015 epoch with the \citet{Hunter:2018} results from VLA observations. \citet{Hunter:2018} used the VLA in "A" configuration and obtained data collected over two sessions on October 29, 2016 and November 19, 2016, overlaid on the ALMA 1 mm continuum map from \citet{Hunter:2017}. They recorded the data in dual polarisation with a channel spacing of 1.953 kHz (0.0878 km~s$^{-1}$) over a span of 90 km~s$^{-1}$ centred on an LSR velocity of $-$7 km~s$^{-1}$. The 6.7-GHz VLA data have an angular resolution approximately a factor of 100 lower than the LBA data, but significantly greater sensitivity. Considering the combined effect of different sensitivities, angular resolution and source evolution, Figure~\ref{fig:maserfeaturesV255ZandHunter2018} shows a strong agreement between the maser distributions. \citet{Hunter:2018} detected many weak maser components (intensities between 0.01 and 0.05 Jy) in MM1 at velocities more positive than $-$4.2 km~s$^{-1}$. We detected some weak emission in this velocity range in the single antenna (autocorrelation) spectra in the October 2015 observations, but this emission was not detected in our VLBI images, showing that, as for most 6.7-GHz methanol maser emission, a significant fraction is spatially resolved out, and indicating that these weak features are not so usually compact. There is a small offset between the LBA and VLA distributions, which has been applied in Figure~\ref{fig:maserfeaturesV255ZandHunter2018} to facilitate better comparison. The VLA observations have accurate absolute astrometry, whereas the LBA observations do not, due to the selection of a poor phase reference source. The VLA distribution is impacted to some degree by the blending of different maser spots within individual maser clusters, and, when combined with the changes over time caused by the flare and the difference in sensitivity, this hinders a direct comparison of the inferred maser distributions. The alignment between the two datasets is hence subjective, and as they have been reduced and imaged using different data reduction packages (AIPS and CASA), projection effects will also impact the relative separation of components at different velocities. The distributions are broadly consistent and in agreement with expectations.
\smallskip

\section{Discussion}\label{Discussion}

NGC6334I is one of the best-studied star formation regions in relation to its maser activity of different species \citep{Weaver:1968}. It hosts multiple maser species in the ground as well as excited states. Slight variability in the 6.7-GHz methanol masers in 1992 and 1993 was reported in NGC6334F \citep{Caswell:1995a}. However, it was unremarkable, so the report on the variability was not included in the first paper investigating the variability of the 6.7-GHz transition \citep{Caswell:1995}. \citet{Goedhart:2004} reported that in February 1999, there was an increase in the 6.7-GHz methanol maser emission at a velocity $-$5.88 km~s$^{-1}$, which peaked in Nov 1999 in the HartRAO's Hartebeesthoek 26m single dish maser monitoring program. The sensitive spectrum of the 6.7-GHz methanol masers undertaken as part of the methanol multibeam survey in 2006/2007 \citep{Caswell:2010} shows the variability to be remarkably similar to the discovery observation by \citet{Menten:1991}.

The detection of a significant flare in a variety of maser transitions associated with NGC6334I in Hartebeesthoek maser monitoring in early 2015 was the catalyst for the LBA observations in October 2015 to enable a comparison with the pre-flare data. It started from 1 Jan 2015 and peaked in August 2015 for the 6.7-GHz class~II methanol masers \citep{MacLeod:2018}. Single dish observations of the 6.7-GHz methanol maser emission in the NGC6334I region before 2015 showed that it covered the velocity range $-12$ to $-6$ km~s$^{-1}$, with the strongest emission towards the most negative velocities \citep[see for example][where NGC6334I corresponds to 351.417$+$0.645 and 351.417$+$0.646]{Caswell:2010}. The 2015 maser flaring event was most noticeable over the velocity range $-$9.0 km~s$^{-1}$ to $-$5 km~s$^{-1}$. The largest changes were found at a velocity of approximately $-7$ km~s$^{-1}$ and the positive limit of the observed emission range expanded to approximately $-4.7$ km~s$^{-1}$ \citep[see figure~1 of][]{MacLeod:2018}. A variety of other follow-up observations were also triggered by the maser flare event, and \citet{Brogan:2016} and \citet{Hunter:2017, Hunter:2018} detected an outburst in the 1.3-mm continuum emission. \citet{Hunter:2018} were able to compare the dust continuum emission at 1.3 mm observed in 2008 with the Submillimeter Array (SMA) with the 6-cm continuum (VLA, 2011) and the 3-mm and 1.3-mm continuum (ALMA, August 2015) \citep{Brogan:2016}, which showed that the NGC6334I protostellar cluster region consists of four primary millimetre sources.

\citet{Hunter:2017} found that dust luminosities of sources in MM1 increased by nearly 70 times between May 2011 and Aug 2015. They reported the location of the outburst in the dust continuum was centred on the continuum source MM1-B and its flux peak is estimated to coincide with the period when \citet{MacLeod:2018} reported an increase in the flux densities in methanol maser in the $-$9.0 km~s$^{-1}$ to $-$5 km~s$^{-1}$ velocity range. Subsequent observations had detected mid-infrared continuum emission post-flare from MM1, indicating an increase in luminosity of approximately 16 times compared to the pre-flare, when MM1 was not detected at 18$\mu$m wavelength \cite{Hunter:2021}. The utility of maser monitoring in detecting and understanding variable mass accretion rates in high-mass star formation regions requires us to understand in much greater detail how and when the two are linked. The large amount of published multi-wavelength data available for the NGC6334I region makes it an ideal source for such investigations. \citet{Hunter:2021}, reported the first mid-IR image of the MM1 post-flare. They found that the outburst in NGC6334I-MM1B exceeded other measured mid-IR outbursts, such as that detected towards S255IR-NIRS3 \citep{CarattioGaratti:2017} by a factor $~\gtrsim~$3 in both duration and energy. They suggest that it requires a hydrodynamic model which involves heating and expansion of the outer layers of the protostar and could explain the longer decay times seen in NGC6334I-MM1 \citep{Hunter:2021, Larson:1980, Herbig:2003}.

It is reasonable to assume that the majority of the observed spectral components arise from the random locations within a volume of molecular gas where there is an unusually high degree of line-of-sight velocity coherence. The presence of interstellar masers implies both suitable physical conditions in the molecular gas (temperature, density, etc), and a sufficient abundance of the relevant molecule with line-of-sight velocity coherence to the observer. Such coherent sight-lines can arise in a variety of ways in massive star-forming regions, including velocity fields associated with expanding HII regions, protostellar outflows, or chance alignments within turbulent molecular gas. In a complex region like NGC6334I, multiple mechanisms may operate simultaneously, making it difficult to identify the origin of any specific maser component. 

\citet{MacLeod:2018} observed a 1-2 orders of magnitude increase in the intensity of several class~II methanol maser components in NGC6334I on timescales of $\sim$6 months, indicating that the excitation conditions changed rapidly across the region. The typical linear extent of class~II methanol maser clusters in the plane of the sky is around 0.03 pc \citep{Caswell:1997}, and there is little evidence of preferred large-scale geometry in NGC6334I. Taken together, these factors suggest that the spatial distribution of maser components is largely random rather than arising from a single organised structure.

The sudden and nearly simultaneous brightening of class~II methanol maser components across MM1 indicates that the excitation conditions changed rapidly over spatial scales of up to $\sim$1000~au from the strongest continuum source MM1-B. Methanol is formed on dust-grain surfaces and released into the gas phase through the desorption of ice-mantles when grains are heated by radiation or shocks \citep{Garrod:2006, Tielens:1997, Hudson:1999, Gibb:1998}. Given that the flare occurred over $\sim$6 months \citep{MacLeod:2018}, the apparent propagation speed of the excitation front is on the order of $\sim$5.6~au/day, corresponding to $\sim$9600~km~s$^{-1}$ (0.03~c) at the source’s distance of 1.3~kpc \citep{WuNGCParallax:2014, Chibueze:2014}. This speed is subluminal, yet far exceeds typical bulk motions of dense molecular gas. Such high apparent speeds cannot be attributed to physical motions of methanol-rich gas. The gas-phase abundance of methanol is known to be relatively low in the presence of shocks exceeding $\sim$10~km~s$^{-1}$ \citep{Garrod:2006, Garay:2002}. Therefore, such rapid morphological changes cannot be due to physical motions of methanol-rich gas. 

Instead, the observations favour a radiatively driven mechanism. In this scenario, enhanced infrared radiation from an accretion outburst at MM1-B heats surrounding dust, liberating methanol into the gas phase and producing maser emission nearly simultaneously across the region. This interpretation is consistent with the subluminal radiative propagation inferred for the high-mass protostellar flare G358-MM1 by \citet{Burns:2020}, who measured apparent expansion rates of 1–2 mas/day (0.04–0.08 c). The coordinated flare across $\sim$1000 au thus supports a radiatively driven excitation mechanism linked to the recent accretion event, rather than shocks or geometrical effects.

The current LBA observations showed the emergence of multiple new clusters of 6.7-GHz class~II methanol maser emission in October 2015. As previously established by \citet{Hunter:2017}, the new 6.7-GHz class~II methanol masers emerged in the molecular gas surrounding the MM1B millimetre continuum source, and \citet{Hunter:2018} reported new maser emission from multiple species in MM1. It is interesting to consider what sort of physical conditions may have changed in and around MM1 that made the conditions suitable for the sudden emergence of the 6.7-GHz methanol, other methanol transitions and maser species in MM1. While changes in methanol abundance, gas temperature/density or motions within the molecular gas can produce maser variability, they are inconsistent with the timescales and changes in intensity that are observed. It is well-established that the class~II methanol masers are radiatively pumped \citep{Cragg:2005, Sobolev:1997} and the light-travel time for the maser cluster regions is comparable to the observed rise-time for the 2015 NGC6334I maser flare. \citet{Hunter:2017} suggested that an unprecedented increase in the temperature of the sub-millimetre dust continuum gave rise to an increase in the radiative pumping in MM1, which then resulted in the emergence of new masers.  \citet{Hunter:2018} suggested that the outburst in the dust continuum heated the dust in the vicinity, which then created the ideal conditions for maser emission inversion in already available methanol-rich gas. The reason for the outburst in the millimetre continuum in the continuum source MM1-B may be an accretion burst. Other possibilities suggested by \citet{Hunter:2018} include a supernova event in MM1 or the merger of two protostars \citep[e.g.][]{BallyZinnecker:2005}.

Based on our high-resolution imaging data, combined with published methanol monitoring and millimetre continuum data, we suggest that in early 2015 an episodic accretion event in one or more protostellar objects in MM1 produced a change in the radiation field in the surrounding molecular gas, which "switched-on" class~II methanol maser emission in this sub-region. The molecular gas within which MM1 is embedded already had a reasonable methanol abundance, as indicated by the 2011 detection of weak thermal emission toward MM1 \citep{Ellingsen:2018}. 

It has previously been suggested that the luminosity of 6.7-GHz methanol masers and the presence of rarer class~II methanol transitions can be used as an indicator of the evolutionary stage of a high-mass star formation region \citep[e.g.][]{Breen:2010, Ellingsen:2011}. In this scenario, sources with low-luminosity 6.7-GHz methanol masers are sources at an early evolutionary stage, and sources like NGC6334I with high-luminosity and detected emission in a plethora of other class~II methanol transitions are the most evolved. The sudden appearance of relatively luminous 6.7-GHz and 12.2-GHz methanol masers, along with the relatively rare 23.1-GHz transition in NGC6334I-MM1, appears to be inconsistent with the suggestion that the methanol masers provide a reliable evolutionary timeline; however, there are a number of potential explanations. One is that for cluster star formation, there will always be objects spanning a range of ages and that the masers are useful only as an evolutionary clock for the protostellar cluster, not individual sources within that cluster. Another is that MM1 in its typical state hosts only relatively low-luminosity 6.7-GHz methanol masers, which were present before 2015, but not detectable in either single-dish or interferometric observations due to dynamic range limitations created by the nearby masers in MM3 and MM2 with overlapping velocity ranges. One relatively easy way to test this would be to make sensitive interferometric observations of class~II methanol transitions such as the 12.2, 19.9, 23.1, 37.7 and 38.3-GHz and compare the data with the published pre-flare distribution of these transitions. Given the absence of any evidence for class~II methanol maser emission associated with MM1 for nearly 25 years, for which we have data before the 2015 flare event, their presence 6 years post-flare would indicate either a permanent shift in the radiation field surrounding MM1, or a very long post-flare decline phase for the masers. The latter has been proposed by \citet{Hunter:2021}, who compared mid-infrared data with the hydrodynamic model of \citet{Meyer:2017} and inferred from them a flare duration of 40-130 years. A repeat of the 2016 millimetre continuum emission observations would also be beneficial, as it would give a more direct and holistic picture of the changes in the radiation field in NGC6334I. The class~II methanol masers associated with both MM2 and MM3 have changed significantly in the period during and post-flare, compared to their behaviour over the preceding $\sim$25 years.  These changes suggest that the 2015 flare has impacted a relatively large volume of the NGC6334I region. The maser emission provides a complementary method of investigating the radiation field. The millimetre continuum observations \citep{Hunter:2018} have angular resolution several orders of magnitude lower than centimetre wavelength VLBI, and their interpretation is complicated by optical depth effects. Maser studies can thus complement millimetre and infrared continuum data by providing information on the locations of the maser flares at much higher resolution, although determining the physical conditions of the masing gas requires observations of multiple transitions and inference in the context of a theoretical model of the masing process.

We have also compared the 6.7-GHz methanol maser emission in NGC6334I, found with the LBA, with that observed by \citet{Ellingsen:2018} in the 37.7-GHz and 38.3-GHz class~II methanol maser transitions (figure 6 in \citet{Ellingsen:2018}). The latter observations were made with the ATCA on March 24, 2011. Figure 6 in \citet{Ellingsen:2018} shows that, for both the 37.7-GHz and 38.3-GHz methanol transitions, maser emission was detected only towards MM2 and MM3. The \citet{Ellingsen:2018} results show weak thermal emission from both transitions, which peaks at a velocity of around $-7$ km s$^{-1}$ towards MM1, but no maser emission. Given that the 37.7/38.3-GHz observations were made approximately 4 years before the NGC6334I flaring event, these data are consistent with the widespread presence of gas-phase methanol in the NGC6334I region, but a lack of widespread inversion of class~II methanol transitions near MM1 before 2015. Similarly, \citet{Krishnan:2013} made ATCA images of the 19.9- and 23.1-GHz methanol maser emission in NGC6334I utilising ATCA observations on March 27, 2005, with those transitions only being detected close to the strongest 6.7-GHz emission from MM3 at that epoch. The 23.1-GHz observations included the velocity range where the strongest class~II methanol maser emission was detected during the 2015 flare, the Hartebeesthoek monitoring of \citet{MacLeod:2018}  shows emission with a peak intensity of around 5~Jy at velocities greater than $-10$ km s$^{-1}$. The 2005 ATCA observations of the 23.1-GHz transition had an RMS noise level of 30 mJy, so we can be confident that any emission from this transition at that time was at least an order of magnitude weaker than in 2015. Sensitive Parkes observations of the 23.1-GHz methanol masers from the NGC6334I region made in 2001 also show no sign of emission at velocities greater than $-10$ km s$^{-1}$ at that epoch \citep{Cragg:2004}. 

\citet{Goedhart:2004} reported a flaring event in the 6.7-GHz methanol masers in NGC6334I, which peaked in November 1999. \citet{Hunter:2018} and \citet{MacLeod:2018} suggested that the 1999 flare and the 2015 flare may originate from the same place in MM1. While we cannot rule out this possibility, there are no interferometric observations before 2015 which show methanol maser emission from MM1, and the velocity range of the 6.7-GHz methanol masers in MM3 extends to $-5.9$~km s$^{-1}$.  There was also an earlier maser flaring event reported by \citet{Weaver:1968} which shows variability in ground-state OH masers in 1965. \citet{MacLeod:2018} have hypothesised that the flaring events in NGC6334I are periodic and predict that the next flaring event may occur in late 2026. However, both interferometric and single dish maser monitoring studies suggest that the earlier maser variability episode reported by \citet{Goedhart:2004} and \citet{Weaver:1968} were less dramatic than the 2015 flaring event, the impact of which is being observed 5 years later, as shown in the March 2020 observations.

The connection between 6.7-GHz methanol maser flares and episodic accretion events in high-mass star formation regions has been established in three cases: S255-IR3, NGC6334I and G358.93-0.03. The first two events provided significant impetus for large-scale monitoring of the 6.7-GHz methanol maser transition. The variability of the class~II methanol masers is typically observed to be less dramatic than for the 22-GHz water masers \citep{Caswell:1995, Ellingsen:2007}, although periodic emission is detected in some sources \citep[e.g.][]{Goedhart:2004}. The reliable identification of maser flaring events triggered by episodic accretion is still being refined, but the key characteristics seem to be a rapid and dramatic increase in the intensity of the 6.7-GHz transition and the appearance of emission from the rarer class~II transitions. These phenomena were observed both in NGC6334I and subsequently in G358.93-0.03 \citep{Breen:2019, Chen:2020, Burns:2020}. G358.93-0.03 is a much lower luminosity 6.7-GHz methanol maser source that showed a dramatic increase in its intensity in a short period in early 2019 \citep{Sugiyama:2019}. The early detection of this flaring event and follow-up coordinated by the Maser Monitoring Organisation (M2O) enabled the detection of a range of rare maser transitions towards G358.93-0.03, many of which were only detectable for a period of weeks \citep{Breen:2019, Chen:2020}. Comparison of targeted mid-infrared observations with archival data confirmed an episodic accretion event in G358.93-0.03, but associated with a less massive protostellar object than NGC6334I-MM1 \citep{Stecklum:2021}. Developing a more detailed understanding of the link between maser variability and episodic accretion events requires both theoretical studies of the chemistry and masers and observational data to inform, constrain and test the theoretical investigations. This work has commenced \citep{Guadarrama:2024}, but is currently in its early stages 

\section{Conclusions}\label{Conclusion}

The unprecedented outburst and maser flaring event in the NGC6334I protostellar cluster presents an excellent opportunity to study the phenomenon of cluster star formation in the Milky Way. Our data are consistent with the maser flare being produced by a sudden change in the radiation field in a region of molecular gas, which already has relatively high methanol abundance. Further high-resolution interferometric observations of methanol and other maser transitions, as well as single dish monitoring observations of the masers and millimetre through mid-infrared continuum observations, will be key to better understanding the processes which govern variable accretion rates in high-mass star-forming regions. 

\section*{Acknowledgements}

This research was supported by the Australian Research Council (ARC) Discovery grants No. DP180101061 and DP230100727. The LBA is part of the Australia Telescope National Facility, which is funded by the Australian Government for operation as a National Facility managed by CSIRO and the University of Tasmania. This research has made use of NASA's Astrophysics Data System Abstract Service. We thank Crystal Brogan (\href{mailto:cbrogan@nrao.edu}{cbrogan@nrao.edu}) and T. R. Hunter (\href{mailto:thunter@nrao.edu}{thunter@nrao.edu}), NRAO, USA, for providing the continuum maps in Figure~\ref{fig:maserfeaturesV255ZandHunter2018}.
\section*{Data Availability}

The correlator FITS files and associated metadata for the Long Baseline Array data underlying this article are available via the Australia Telescope Online Archive under experiment code V255 sessions I, J, Z and AI.



\bibliographystyle{mnras}
\bibliography{JayReference} 




\appendix

\section{Details of All Maser Features for NGC6334I Emission for All Epochs}\label{appendixA:ngc6334bigtables}

\onecolumn
\begin{longtable}{c|l|c|c|c|c}
\caption{Information for all the maser features found in NGC6334I in March 2010. Columns 1--2 list the feature ID number and its association with a particular sub-region; columns 3-6 list the LSR velocity, R.A., DEC. and integrated flux of the feature.}\label{tab:Detailsv255Iallmaserfeatures}\\
\toprule
\hline
\multicolumn{1}{c}{\bf Feature} &
\multicolumn{1}{c}{\bf feature}  &
\multicolumn{1}{c}{\bf Velocity}  &
\multicolumn{1}{c}{\bf R. A. Offset}   &
\multicolumn{1}{c}{\bf DEC. Offset}   &
\multicolumn{1}{c}{\bf Integrated Flux}  \\
\multicolumn{1}{c}{\bf Number} &
\multicolumn{1}{c}{\bf Association}  &
\multicolumn{1}{c}{\bf (kms$^{-1}$)}  &
\multicolumn{1}{c}{\bf (mas)}   &
\multicolumn{1}{c}{\bf (mas)}   &
\multicolumn{1}{c}{\bf (Jy~beam$^{-1}$)}  \\
\toprule
\hline
1   &  MM2  &  -11.499    &      	-2472.826 	&   2227.456	   &   5.81           \\
2   &  MM2  &  -11.473    &      	-2533.170 	&   2159.186	   &   3.35           \\
3   &  MM2  &  -11.374    &      	-2364.274 	&   2155.728	   &  11.80           \\
4   &  MM2  &  -11.364    &      	-2503.921 	&   1930.274	   &   9.11           \\
5   &  MM2  &  -11.350    &      	-2287.366 	&   1845.490	   &   9.23           \\
6   &  MM2  &  -11.322    &      	-2373.798 	&   2156.358	   &  14.40           \\
7   &  MM2  &  -11.319    &      	-2530.383 	&   2195.336	   &  22.10           \\
8   &  MM2  &  -11.298    &      	-2376.946 	&   2330.817	   &  10.20           \\
9   &  MM2  &  -11.276    &      	-2360.811 	&   2250.616	   &  17.00           \\
10  &  MM2  &  -11.227    &      	-2389.125 	&   2153.083	   &  25.30           \\
11  &  MM2  &  -11.188    &      	-2411.281 	&   2327.033	   &  14.20           \\
12  &  MM2  &  -11.177    &      	-2367.577 	&   2246.295	   & 233.00           \\
13  &  MM2  &  -11.172    &      	-2407.694 	&   2149.681	   &  16.90           \\
14  &  MM2  &  -11.144    &      	-2426.028 	&   2411.946	   &  15.30           \\
15  &  MM2  &  -11.129    &      	-2405.204 	&   2238.906	   &  50.20           \\
16  &  MM2  &  -11.123    &      	-2391.684 	&   2239.600	   &  20.70           \\
17  &  MM2  &  -11.078    &      	-2420.727 	&   2148.180	   &  11.40           \\
18  &  MM2  &  -11.042    &      	-2421.225 	&   2231.044	   &  12.20           \\
19  &  MM2  &  -11.034    &      	-2428.747 	&   2235.703	   &  16.30           \\
20  &  MM2  &  -10.998    &      	-2408.222 	&   2238.077	   &  32.40           \\
21  &  MM2  &  -10.942    &      	-2400.393 	&   2242.556	   &  21.50           \\
22  &  MM2  &  -10.915    &      	-2448.615 	&   2237.112	   &  13.40           \\
23  &  MM2  &  -10.844    &      	-2385.235 	&   2335.021	   &   3.74           \\
24  &  MM3  &  -10.771    &           -73.272 	&   -485.852	   &  80.20           \\
25  &  MM3  &  -10.749    &           876.527 	&    478.920	   &   5.73           \\
26  &  MM2  &  -10.744    &      	-2461.727 	&   2239.474	   &  17.50           \\
27  &  MM3  &  -10.654    &      	  -67.876 	&   -200.658	   &  12.20           \\
28  &  MM3  &  -10.639    &      	  -46.726 	&   -106.601	   & 113.00           \\
29  &  MM3  &  -10.610    &      	  -50.873 	&   -103.580	   &  12.50           \\
30  &  MM3  &  -10.599    &      	  -56.146 	&   -105.408	   & 145.00           \\
31  &  MM3  &  -10.358    &      	  -43.949 	&    -75.895	   &  35.30           \\
32  &  MM3  &  -10.332    &      	  -17.438 	&      8.495	   &  16.70           \\
33  &  MM3  &  -10.288    &      	  -45.275 	&    -81.271	   &  58.20           \\
34  &  MM3  &  -10.266    &      	  -56.504 	&    -66.210	   &  51.90           \\
35  &  MM3  &  -10.244    &      	  -36.258 	&      4.511	   &  40.50           \\
36  &  MM3  &  -10.244    &      	  -58.668 	&    -77.856	   &  54.60           \\
37  &  MM3  &  -10.238    &      	  -22.684 	&      9.593	   &  59.60           \\
38  &  MM3  &  -10.221    &      	   -9.581 	&     16.556	   &  24.00           \\
39  &  MM2  &  -10.200    &      	-2568.260 	&   2223.409	   &  17.30           \\
40  &  MM3  &  -10.182    &      	   -9.973 	&     10.290	   & 271.00           \\
41  &  MM3  &  -10.135    &      	   -8.376 	&   -245.085	   &   6.22           \\
42  &  MM2  &  -10.127    &      	-2539.426 	&   2232.296	   &   9.24           \\
43  &  MM2  &  -10.105    &      	-2625.868 	&   2152.789	   &  11.50           \\
44  &  MM3  &  -10.069    &      	  -10.969 	&     -3.971	   &  11.10           \\
45  &  MM2  &  -10.069    &      	-2564.677 	&   2223.636	   &  13.00           \\
46  &  MM2  &  -10.031    &      	-2632.512 	&   2241.019	   &   5.92           \\
47  &  MM2  &  -10.019    &      	-2621.664 	&   2240.601	   &   5.66           \\
48  &  MM2  &  -10.003    &      	-2584.245 	&   2315.909	   &   6.11           \\
49  &  MM2  &  -10.003    &      	-2573.242 	&   2314.089	   &   8.29           \\
50  &  MM2  &   -9.985    &         -2568.721 	&   2229.057	   &   7.47           \\
51  &  MM2  &   -9.981    &         -2547.810 	&   2318.327	   &   9.09           \\
52  &  MM2  &   -9.959    &         -2608.242 	&   2246.833	   &   3.14           \\
53  &  MM2  &   -9.959    &         -2596.362 	&   2245.858	   &   3.57           \\
54  &  MM2  &   -9.928    &         -2549.129 	&   2229.228	   &   7.12           \\
55  &  MM3  &   -9.824    &           -23.679 	&    -88.130	   &   5.31           \\
56  &  MM3  &   -9.805    &           -12.933 	&      5.459	   &   2.82           \\
57  &  MM3  &   -9.783    &            83.555 	&    -56.283	   &  48.90           \\
58  &  MM3  &   -9.761    &             0.858 	&      4.333	   &   5.04           \\
59  &  MM3  &   -9.723    &            -1.987 	&     -4.956	   &  23.70           \\
60  &  MM3  &   -9.698    &             7.104 	&     -1.918	   &   9.42           \\
61  &  MM3  &   -9.652    &          -165.251 	&    -49.995	   &  10.40           \\
62  &  MM3  &   -9.622    &            -6.394 	&      1.493	   &   4.54           \\
63  &  MM3  &   -9.612    &            -0.042 	&     -0.158	   & 129.00           \\
64  &  MM3  &   -9.542    &           165.025 	&     50.332	   &  48.70           \\
65  &  MM3  &   -9.527    &           172.331 	&     46.660	   &   3.69           \\
66  &  MM2  &   -9.322    &         -2764.427 	&   2535.737	   &   2.34           \\
67  &  MM3  &   -9.235    &           548.979 	&  -1824.108	   &   2.04           \\
68  &  MM2  &   -9.037    &         -2748.269 	&   2262.438	   &   6.68           \\
69  &  MM3  &   -9.037    &           721.713 	&    -54.553	   &   2.23           \\
70  &  MM2  &   -9.008    &         -2754.765 	&   2266.170	   &   2.98           \\
71  &  MM3  &   -8.971    &           481.087 	&     23.836	   &   1.28           \\
72  &  MM3  &   -8.927    &           140.087 	&     45.764	   &   7.00           \\
73  &  MM3  &   -8.884    &           133.784 	&    -14.718	   &   1.06           \\
74  &  MM3  &   -8.883    &           126.392 	&    -11.667	   &  11.60           \\
75  &  MM3  &   -8.730    &           708.730 	&    -96.158	   &   6.67           \\
76  &  MM3  &   -8.708    &           746.185 	&   -124.969	   &   1.07           \\
77  &  MM3  &   -8.708    &           705.839 	&   -101.074	   &   7.54           \\
78  &  MM3  &   -8.576    &           737.237 	&   -120.902	   &   4.30           \\
79  &  MM3  &   -8.510    &           567.807 	&   -175.763	   &   8.35           \\
80  &  MM3  &   -8.488    &           206.012 	&    862.979	   &   2.17           \\
81  &  MM3  &   -8.488    &           596.831 	&    994.523	   &   2.18           \\
82  &  MM3  &   -8.482    &           733.431 	&   -125.347	   &  59.90           \\
83  &  MM3  &   -8.313    &           464.357 	&  -1220.254	   &   3.55           \\
84  &  MM3  &   -8.292    &           740.116 	&   -130.138	   &   5.20           \\
85  &  MM3  &   -8.181    &           452.853 	&     61.062	   &   1.23           \\
86  &  MM3  &   -8.005    &           471.490 	&  -1223.624	   &   3.85           \\
87  &  MM3  &   -7.930    &           746.915 	&     49.153	   &   0.62           \\
88  &  MM3  &   -7.874    &           750.336 	&   -125.398	   &   7.57           \\
89  &  MM2  &   -7.742    &         -2976.020 	&   2870.709	   &   7.01           \\
90  &  MM3  &   -7.413    &           509.682 	&     85.966	   &   5.02           \\
91  &  MM3  &   -7.369    &           347.900 	&  -2023.312	   &   2.10           \\
92  &  MM3  &   -7.347    &           513.082 	&  -1972.761	   &  23.10           \\
93  &  MM3  &   -7.347    &           522.160 	&  -1977.426	   &   2.16           \\
94  &  MM3  &   -7.186    &           507.840 	&     94.452	   &   0.76           \\
95  &  MM3  &   -7.184    &           589.855 	&  -1969.865	   &   2.02           \\
96  &  MM3  &   -7.106    &           629.657 	&  -1987.981	   &   0.95           \\
97  &  MM3  &   -7.047    &           515.404 	&     92.411	   &   2.72           \\
98  &  MM3  &   -7.040    &           520.813 	&  -1994.356       &   1.67           \\
99  &  MM3  &   -6.933    &           615.061 	&  -1981.326       &  10.50           \\
100 &  MM3  &   -6.864    &           659.708 	&  -2020.389       &   1.59           \\
101 &  MM3  &   -6.864    &           684.563 	&  -1918.393       &   0.99           \\
102 &  MM3  &   -6.842    &           648.407 	&  -2020.035       &   2.48           \\
103 &  MM3  &   -6.820    &           668.716 	&  -2029.311       &   1.02           \\
104 &  MM3  &   -6.738    &           610.923 	&  -1979.441       &  10.10           \\
105 &  MM3  &   -6.381    &           650.804 	&  -2026.798       &   2.60           \\
106 &  MM3  &   -6.337    &           686.136 	&  -2028.616       &   0.94           \\
\hline
\bottomrule
\end{longtable}


\begin{longtable}{c|l|c|c|c|c}
\caption{Same as Table~\ref{tab:Detailsv255Iallmaserfeatures}, but in July 2010.}\label{tab:Detailsv255Jallmaserfeatures}\\
\toprule
\hline
\multicolumn{1}{c}{\bf Feature} &
\multicolumn{1}{c}{\bf feature}  &
\multicolumn{1}{c}{\bf Velocity}  &
\multicolumn{1}{c}{\bf R. A. Offset}   &
\multicolumn{1}{c}{\bf DEC. Offset}   &
\multicolumn{1}{c}{\bf Integrated Flux}  \\
\multicolumn{1}{c}{\bf Number} &
\multicolumn{1}{c}{\bf Association}  &
\multicolumn{1}{c}{\bf (kms$^{-1}$)}  &
\multicolumn{1}{c}{\bf (mas)}   &
\multicolumn{1}{c}{\bf (mas)}   &
\multicolumn{1}{c}{\bf (Jy~beam$^{-1}$)}  \\
\toprule
\hline
1  & MM2  & -11.612    &    -2483.033    &       2229.044    &  0.66    \\
2  & MM2  & -11.524    &    -2304.709    &       2286.687    &  3.65    \\
3  & MM2  & -11.524    &    -2293.835    &       2279.895    &  2.73    \\
4  & MM2  & -11.502    &    -2589.443    &       2078.522    &  3.09    \\
5  & MM2  & -11.502    &    -2417.019    &       2172.457    &  9.49    \\
6  & MM2  & -11.491    &    -2350.971    &       2242.419    &  4.24    \\
7  & MM2  & -11.460    &    -2372.650    &       2215.655    &  34.80   \\
8  & MM2  & -11.458    &    -2507.389    &       2022.525    &  7.73    \\
9  & MM2  & -11.436    &    -2314.547    &       2294.818    &  7.11    \\
10 & MM2  & -11.421    &    -2388.987    &       2162.437    &  7.39    \\
11 & MM2  & -11.414    &    -2350.048    &       2475.445    &  4.98    \\
12 & MM2  & -11.392    &    -2326.460    &       2476.380    &  5.92    \\
13 & MM2  & -11.392    &    -2332.373    &       2475.943    &  6.47    \\
14 & MM2  & -11.387    &    -2552.168    &       2165.329    &  18.00   \\
15 & MM2  & -11.378    &    -2324.984    &       2298.986    &  41.40   \\
16 & MM2  & -11.370    &    -2211.950    &       2195.727    &  10.70   \\
17 & MM2  & -11.370    &    -2310.846    &       2288.095    &  12.50   \\
18 & MM2  & -11.370    &    -1589.251    &       2273.446    &  4.82    \\
19 & MM2  & -11.370    &    -2179.394    &       2306.951    &  5.79    \\
20 & MM2  & -11.370    &    -2295.238    &       2560.587    &  7.68    \\
21 & MM2  & -11.355    &    -2403.610    &       2165.393    &  9.34    \\
22 & MM2  & -11.348    &    -2535.359    &       1934.436    &  13.10   \\
23 & MM2  & -11.348    &    -2093.715    &       2214.490    &  8.51    \\
24 & MM2  & -11.346    &    -2373.293    &       2249.887    &  10.60   \\
25 & MM2  & -11.343    &    -2162.359    &       2284.364    &  7.31    \\
26 & MM2  & -11.334    &    -2385.821    &       2223.406    &  16.40   \\
27 & MM2  & -11.326    &    -2196.117    &       2315.070    &  9.76    \\
28 & MM2  & -11.326    &    -1885.950    &       2145.142    &  10.10   \\
29 & MM2  & -11.282    &    -2221.354    &       2196.895    &  12.80   \\
30 & MM2  & -11.282    &    -2503.037    &       2247.887    &  18.90   \\
31 & MM2  & -11.282    &    -2320.323    &       2794.691    &  7.61    \\
32 & MM2  & -11.260    &    -2358.285    &       2244.301    &  64.40   \\
33 & MM2  & -11.254    &    -2380.833    &       2215.925    &  22.00   \\
34 & MM2  & -11.249    &    -2496.202    &       2010.281    &  15.30   \\
35 & MM2  & -11.238    &    -2346.778    &       2300.698    &  7.10    \\
36 & MM2  & -11.222    &    -2565.134    &       2165.614    &  21.20   \\
37 & MM2  & -11.216    &    -2172.016    &       2290.199    &  8.72    \\
38 & MM2  & -11.216    &    -2305.073    &       2281.189    &  14.60   \\
39 & MM2  & -11.198    &    -2366.696    &       2245.056    &  99.00   \\
40 & MM2  & -11.194    &    -2333.929    &       2297.565    &  35.10   \\
41 & MM2  & -11.129    &    -2527.086    &       1928.917    &  11.30   \\
42 & MM2  & -11.128    &    -2409.194    &       2240.078    &  29.80   \\
43 & MM2  & -11.107    &    -2554.381    &       1935.026    &  8.09    \\
44 & MM2  & -11.092    &    -2312.155    &       2282.436    &  9.26    \\
45 & MM2  & -11.078    &    -2394.510    &       2157.296    &  23.50   \\
46 & MM2  & -10.975    &    -2551.672    &       1842.825    &  5.07    \\
47 & MM2  & -10.953    &    -2405.354    &       2236.828    &  8.95    \\
48 & MM2  & -10.909    &    -2386.977    &       2153.452    &  3.37    \\
49 & MM2  & -10.821    &    -2454.093    &       2238.702    &  10.20   \\
50 & MM3  & -10.821    &    -527.155     &      -54.799      &  3.48    \\
51 & MM3  & -10.799    &    -74.087      &      -486.639     &  43.20   \\
52 & MM3  & -10.799    &    -82.727      &      -515.962     &  3.91    \\
53 & MM3  & -10.799    &    -126.551     &      -558.892     &  4.94    \\
54 & MM3  & -10.777    &    -93.283      &      -510.250     &  14.00   \\
55 & MM3  & -10.777    &    -271.860     &      -566.205     &  5.53    \\
56 & MM3  & -10.755    &    -66.691      &      -485.575     &  30.00   \\
57 & MM3  & -10.755    &    -39.656      &      -111.133     &  10.40   \\
58 & MM3  & -10.746    &    -265.399     &      -562.751     &  5.94    \\
59 & MM2  & -10.744    &    -2462.871    &       2238.691    &  6.78    \\
60 & MM3  & -10.740    &    -13.271      &      -452.378     &  4.43    \\
61 & MM3  & -10.733    &    121.278      &      -442.505     &  7.31    \\
62 & MM3  & -10.720    &    -273.895     &      -559.159     &  6.45    \\
63 & MM3  & -10.711    &     28.291      &      -93.133      &  7.04    \\
64 & MM3  & -10.668    &    -94.950      &      -187.475     &  15.50   \\
65 & MM3  & -10.646    &    -3.555       &      -62.288      &  7.99    \\
66 & MM3  & -10.624    &    -106.942     &      -182.333     &  16.40   \\
67 & MM3  & -10.624    &    -13.226      &      -54.373      &  18.10   \\
68 & MM3  & -10.623    &    -49.501      &      -105.312     &  179.00  \\
69 & MM3  & -10.580    &    -57.846      &      -104.174     &  24.80   \\
70 & MM3  & -10.404    &    -54.201      &      -96.159      &  32.90   \\
71 & MM3  & -10.404    &    -220.641     &      -62.880      &  20.90   \\
72 & MM3  & -10.404    &     25.218      &       471.370     &  12.40   \\
73 & MM3  & -10.382    &     30.030      &      -24.130      &  20.10   \\
74 & MM3  & -10.382    &    -151.123     &       13.762      &  26.20   \\
75 & MM3  & -10.360    &    -210.211     &      -69.222      &  30.40   \\
76 & MM3  & -10.360    &     29.725      &       64.855      &  16.70   \\
77 & MM3  & -10.360    &    -72.113      &      -66.067      &  22.50   \\
78 & MM3  & -10.347    &    -17.865      &       7.102       &  30.90   \\
79 & MM3  & -10.316    &     37.186      &      -29.420      &  16.60   \\
80 & MM3  & -10.279    &    -56.734      &      -72.788      &  43.40   \\
81 & MM3  & -10.272    &    -64.071      &      -70.685      &  74.80   \\
82 & MM3  & -10.258    &    -34.647      &       3.075       &  13.20   \\
83 & MM3  & -10.251    &    -29.172      &      -4.113       &  20.10   \\
84 & MM3  & -10.250    &    -52.133      &      -76.583      &  64.10   \\
85 & MM3  & -10.218    &    -15.117      &      -11.368      &  11.00   \\
86 & MM3  & -10.214    &    -147.516     &       13.122      &  8.98    \\
87 & MM2  & -10.185    &    -2577.534    &       2228.728    &  14.60   \\
88 & MM3  & -10.167    &    -19.462      &       14.976      &  25.60   \\
89 & MM3  & -10.163    &    -215.106     &      -63.034      &  10.70   \\
90 & MM3  & -10.162    &    -11.763      &       10.053      &  196.00  \\
91 & MM3  & -10.145    &    -7.731       &       16.544      &  13.70   \\
92 & MM3  & -10.141    &     36.519      &      -164.373     &  9.14    \\
93 & MM2  & -10.116    &    -2635.195    &       2241.549    &  14.50   \\
94 & MM3  & -10.075    &    -11.298      &      -3.958       &  14.60   \\
95 & MM2  & -10.075    &    -2574.187    &       2228.874    &  10.30   \\
96 & MM3  & -10.024    &    -6.103       &       7.467       &  96.40   \\
97 & MM2  & -10.009    &    -2583.532    &       2234.325    &  9.01    \\
98 & MM2  &  -9.993    &    -2639.296    &     	 2245.427    &  12.00   \\
99 & MM2  &  -9.987    &    -2570.870    &     	 2229.753    &  6.03    \\
100 & MM2 &  -9.943    &    -2548.925    &       2228.882    &  6.62    \\
101 & MM2 &  -9.899    &    -2567.459    &       2231.090    &  3.23    \\
102 & MM3 &  -9.789    &     83.524      &      -56.317      &  52.40   \\
103 & MM3 &  -9.775    &     115.587     &       30.851      &  5.97    \\
104 & MM3 &  -9.724    &     550.991     &      -763.218     &  4.08    \\
105 & MM3 &  -9.713    &     6.566       &      -3.655       &  25.00   \\
106 & MM3 &  -9.702    &     659.902     &       1478.524    &  3.96    \\
107 & MM3 &  -9.658    &    -165.568     &      -50.029      &  8.20    \\
108 & MM3 &  -9.635    &     0.002       &      -0.163       &  129.00  \\
109 & MM3 &  -9.526    &     165.225     &       50.395      &  50.30   \\
110 & MM3 &  -9.526    &     322.401     &       88.415      &  3.26    \\
111 & MM3 &  -9.515    &     171.785     &       46.208      &  3.60    \\
112 & MM3 &  -9.504    &    -11.827      &       6.406       &  2.21    \\
113 & MM2 &  -9.087    &    -2781.559    &       2210.767    &  1.62    \\
114 & MM2 &  -9.065    &    -2747.454    &       2262.187    &  7.73    \\
115 & MM3 &  -8.955    &     140.409     &       45.950      &  5.65    \\
116 & MM3 &  -8.911    &     119.175     &      -8.995       &  1.35    \\
117 & MM3 &  -8.889    &     126.802     &      -11.318      &  10.30   \\
118 & MM3 &  -8.771    &     706.314     &      -101.087     &  5.62    \\
119 & MM3 &  -8.747    &     709.583     &      -95.880      &  4.55    \\
120 & MM3 &  -8.736    &     700.806     &      -94.959      &  1.37    \\
121 & MM3 &  -8.648    &     738.578     &      -122.292     &  2.75    \\
122 & MM3 &  -8.648    &     942.685     &      -48.934      &  2.59    \\
123 & MM3 &  -8.582    &     517.530     &      -265.205     &  2.97    \\
124 & MM3 &  -8.560    &     564.125     &      -182.492     &  3.45    \\
125 & MM3 &  -8.545    &     482.162     &      -121.036     &  2.05    \\
126 & MM3 &  -8.538    &     733.955     &      -125.197     &  45.90   \\
127 & MM3 &  -8.538    &     555.314     &      -177.018     &  4.25    \\
128 & MM3 &  -8.538    &     865.346     &      -112.273     &  3.78    \\
129 & MM3 &  -8.538    &     891.861     &      -87.127      &  3.43    \\
130 & MM3 &  -8.538    &     839.481     &       849.494     &  2.53    \\
131 & MM3 &  -8.516    &     890.679     &      -145.606     &  2.39    \\
132 & MM3 &  -8.516    &     562.741     &      -406.988     &  2.17    \\
133 & MM3 &  -8.516    &     462.665     &      -92.019      &  2.34    \\
134 & MM3 &  -8.501    &     683.692     &       854.129     &  2.61    \\
135 & MM3 &  -8.501    &    -821.750     &       837.544     &  2.03    \\
136 & MM3 &  -8.494    &     756.212     &      -98.650      &  2.44    \\
137 & MM3 &  -8.494    &     833.326     &       536.102     &  2.03    \\
138 & MM3 &  -8.485    &     753.335     &      -96.307      &  5.24    \\
139 & MM3 &  -8.450    &     687.510     &      -169.158     &  3.13    \\
140 & MM3 &  -8.428    &     725.150     &      -125.322     &  3.56    \\
141 & MM3 &  -8.428    &     888.647     &       519.780     &  1.91    \\
142 & MM3 &  -8.406    &     698.032     &      -174.688     &  5.07    \\
143 & MM3 &  -8.406    &     702.392     &      -179.587     &  4.53    \\
144 & MM3 &  -8.377    &     708.205     &      -178.992     &  6.14    \\
145 & MM3 &  -8.350    &     761.657     &      -104.024     &  4.12    \\
146 & MM3 &  -8.341    &     458.691     &      -1216.745    &  1.14    \\
147 & MM3 &  -8.340    &     741.771     &      -130.175     &  5.41    \\
148 & MM3 &  -8.209    &     735.926     &      -131.509     &  2.09    \\
149 & MM3 &  -8.129    &     713.948     &      -184.913     &  1.72    \\
150 & MM3 &  -8.121    &     469.746     &      -1222.001    &  2.41    \\
151 & MM3 &  -8.008    &     491.107     &      -1193.994    &  0.71    \\
152 & MM3 &  -7.960    &     768.211     &      -98.673      &  1.18    \\
153 & MM3 &  -7.887    &     771.743     &      -98.836      &  1.17    \\
154 & MM3 &  -7.858    &     721.106     &      -179.410     &  1.60    \\
155 & MM3 &  -7.858    &     752.133     &      -124.558     &  3.35    \\
156 & MM2 &  -7.836    &    -3155.626    &       2816.832    &  0.68    \\
157 & MM2 &  -7.792    &    -3164.651    &       2819.426    &  0.65    \\
158 & MM2 &  -7.792    &    -2975.745    &       2870.974    &  2.84    \\
159 & MM2 &  -7.792    &    -3006.119    &       2815.767    &  2.09    \\
160 & MM2 &  -7.770    &    -2956.270    &       2897.990    &  2.58    \\
161 & MM3 &  -7.399    &     530.910     &       111.718     &  1.65    \\
162 & MM3 &  -7.397    &     481.118     &      -2023.966    &  6.97    \\
163 & MM3 &  -7.384    &     505.100     &      -1972.919    &  1.83    \\
164 & MM3 &  -7.375    &     324.547     &      -2023.828    &  1.32    \\
165 & MM3 &  -7.375    &     514.267     &      -1972.584    &  5.12    \\
166 & MM3 &  -7.375    &     532.311     &      -1945.534    &  10.00   \\
167 & MM3 &  -7.375    &     314.094     &      -2018.139    &  1.41    \\
168 & MM3 &  -7.366    &     584.392     &      -1991.586    &  1.03    \\
169 & MM3 &  -7.353    &     394.083     &      -1946.882    &  1.49    \\
170 & MM3 &  -7.353    &     646.788     &      -1959.669    &  2.22    \\
171 & MM3 &  -7.353    &     1024.234    &       661.432     &  0.86    \\
172 & MM3 &  -7.309    &     473.362     &      -2026.819    &  2.89    \\
173 & MM3 &  -7.257    &     464.117     &      -2022.744    &  1.19    \\
174 & MM3 &  -7.140    &     548.530     &      -1969.070    &  1.04    \\
175 & MM3 &  -7.089    &     661.015     &      -1982.576    &  1.09    \\
176 & MM3 &  -7.008    &     626.031     &      -1951.703    &  3.69    \\
177 & MM3 &  -7.002    &     641.033     &      -1961.103    &  1.36    \\
178 & MM3 &  -6.936    &     588.100     &      -2034.767    &  2.00    \\
179 & MM3 &  -6.924    &     561.406     &      -2023.097    &  1.85    \\
180 & MM3 &  -6.914    &     500.024     &      -1955.963    &  1.42    \\
181 & MM3 &  -6.914    &     575.591     &      -2033.511    &  4.70    \\
182 & MM3 &  -6.913    &     630.946     &      -1954.829    &  10.20   \\
183 & MM3 &  -6.892    &     619.028     &      -1980.921    &  1.25    \\
184 & MM3 &  -6.892    &     613.488     &      -1950.106    &  1.79    \\
185 & MM3 &  -6.881    &     580.590     &      -1893.288    &  1.31    \\
186 & MM3 &  -6.870    &     623.980     &      -1984.212    &  1.24    \\
187 & MM3 &  -6.848    &     755.617     &      -1821.997    &  1.02    \\
188 & MM3 &  -6.584    &     773.473     &      -1946.977    &  3.52    \\
\hline
\bottomrule
\end{longtable}


\begin{longtable}{c|l|c|c|c|c}
\caption{Same as Table~\ref{tab:Detailsv255Iallmaserfeatures}, but in October 2015.}\label{tab:Detailsv255Zallmaserfeatures}\\
\toprule
\hline
\multicolumn{1}{c}{\bf Feature} &
\multicolumn{1}{c}{\bf feature}  &
\multicolumn{1}{c}{\bf Velocity}  &
\multicolumn{1}{c}{\bf R. A. Offset}   &
\multicolumn{1}{c}{\bf DEC. Offset}   &
\multicolumn{1}{c}{\bf Integrated Flux}  \\
\multicolumn{1}{c}{\bf Number} &
\multicolumn{1}{c}{\bf Association}  &
\multicolumn{1}{c}{\bf (kms$^{-1}$)}  &
\multicolumn{1}{c}{\bf (mas)}   &
\multicolumn{1}{c}{\bf (mas)}   &
\multicolumn{1}{c}{\bf (Jy~beam$^{-1}$)}  \\
\toprule
\hline
1    &  MM2       &      -11.426 &   -2354.789  &   2243.027     &   14.70   \\
2    &  MM2       &      -11.426 &   -2363.506  &   2236.303     &   4.17    \\
3    &  MM2       &      -11.404 &   -2550.967  &   2435.696     &   3.58    \\
4    &  MM2       &      -11.294 &   -2363.248  &   2244.380     &   60.00   \\
5    &  MM2       &      -11.292 &   -2370.528  &   2238.521     &   28.00   \\
6    &  MM2       &      -11.287 &   -2383.166  &   1815.333     &   4.57    \\
7    &  MM2       &      -11.273 &   -2374.940  &   2244.950     &   4.21    \\
8    &  MM2       &      -11.229 &   -2075.626  &   1904.068     &   5.77    \\
9    &  MM2       &      -11.229 &   -2382.101  &   2249.409     &   12.50   \\
10   &  MM2       &      -11.229 &   -2315.520  &   2759.465     &   7.09    \\
11   &  MM2       &      -11.229 &   -2567.145  &   2438.779     &   9.49    \\
12   &  MM2       &      -11.204 &   -2387.640  &   2235.926     &   35.80   \\
13   &  MM2       &      -11.189 &   -2380.618  &   2238.634     &   40.40   \\
14   &  MM2       &      -11.163 &   -2322.387  &   2759.817     &   5.10    \\
15   &  MM2       &      -11.163 &   -2403.871  &   2232.345     &   8.72    \\
16   &  MM2       &      -11.155 &   -2386.556  &   2232.271     &   11.10   \\
17   &  MM2       &      -11.152 &   -2577.066  &   2433.860     &   11.70   \\
18   &  MM2       &      -11.119 &   -2388.116  &   2244.830     &   8.60    \\
19   &  MM2       &      -11.097 &   -2421.032  &   2229.890     &   6.09    \\
20   &  MM2       &      -11.097 &   -2327.781  &   2760.588     &   6.85    \\
21   &  MM2       &      -11.097 &   -2395.108  &   2241.598     &   5.66    \\
22   &  MM2       &      -11.082 &   -2634.310  &   3095.808     &   5.43    \\
23   &  MM2       &      -11.078 &   -2400.485  &   2234.624     &   22.60   \\
24   &  MM2       &      -11.075 &   -2394.089  &   2236.093     &   57.10   \\
25   &  MM2       &      -11.053 &   -2581.870  &   2438.181     &   17.40   \\
26   &  MM2       &      -11.053 &   -1905.723  &   1706.482     &   9.03    \\
27   &  MM2       &      -11.009 &   -2413.954  &   2231.341     &   11.90   \\
28   &  MM2       &      -11.002 &   -2402.058  &   2246.599     &   20.80   \\
29   &  MM2       &      -10.979 &   -2454.345  &   2900.405     &   5.88    \\
30   &  MM2       &      -10.979 &   -2590.686  &   2436.132     &   7.21    \\
31   &  MM2       &      -10.932 &   -2406.995  &   2236.318     &   31.60   \\
32   &  MM2       &      -10.760 &   -2462.030  &   2237.478     &   3.84    \\
33   &  MM3       &      -10.746 &   -73.237    &  -487.331      &   40.30   \\
34   &  MM3       &      -10.658 &   -48.426    &  -109.103      &   21.00   \\
35   &  MM3       &      -10.636 &   -56.551    &  -108.153      &   40.50   \\
36   &  MM3       &      -10.614 &   -358.652   &   227.517      &   13.80   \\
37   &  MM3       &      -10.599 &   -63.295    &  -99.783       &   36.70   \\
38   &  MM3       &      -10.584 &   -359.891   &   233.431      &   13.80   \\
39   &  MM3       &      -10.570 &   -56.765    &  -101.980      &   63.70   \\
40   &  MM3       &      -10.438 &    243.621   &  -424.067      &   29.40   \\
41   &  MM3       &      -10.409 &   -69.456    &  -87.814       &   24.60   \\
42   &  MM3       &      -10.372 &   -3.051     &   9.231        &   20.20   \\
43   &  MM3       &      -10.372 &    237.458   &  -414.362      &   13.60   \\
44   &  MM3       &      -10.307 &    253.197   &  -414.353      &   15.00   \\
45   &  MM3       &      -10.307 &    244.653   &  -412.587      &   25.20   \\
46   &  MM3       &      -10.306 &    55.317    &  -217.090      &   55.10   \\
47   &  MM3       &      -10.306 &   -247.745   &   117.621      &   18.30   \\
48   &  MM3       &      -10.277 &    165.942   &  -183.131      &   16.00   \\
49   &  MM3       &      -10.263 &   -246.090   &   120.439      &   23.70   \\
50   &  MM3       &      -10.263 &   -431.615   &   314.814      &   25.40   \\
51   &  MM3       &      -10.262 &   -180.054   &   639.767      &   38.00   \\
52   &  MM3       &      -10.248 &   -51.867    &  -249.688      &   13.40   \\
53   &  MM3       &      -10.241 &   -13.555    &   8.744        &   41.60   \\
54   &  MM3       &      -10.219 &   -57.295    &  -76.572       &   30.00   \\
55   &  MM3       &      -10.219 &    65.078    &  -218.548      &   19.00   \\
56   &  MM3       &      -10.183 &   -6.388     &  -0.434        &   16.00   \\
57   &  MM3       &      -10.183 &   -20.114    &   18.990       &   13.70   \\
58   &  MM3       &      -10.169 &   -9.577     &   11.075       &   361.0   \\
59   &  MM3       &      -10.160 &   -427.565   &   315.094      &   12.90   \\
60   &  MM3       &      -10.021 &   -11.080    &  -3.499        &   62.10   \\
61   &  MM2       &      -9.880  &   -2391.304  &   2266.745     &   5.30    \\
62   &  MM3       &      -9.824  &    83.397    &  -56.046       &   15.90   \\
63   &  MM3       &      -9.685  &   -0.042     &  -0.138        &   280.0   \\
64   &  MM3       &      -9.494  &    164.768   &   50.536       &   70.60   \\
65   &  MM1-II    &      -9.213  &   -441.178   &   2920.425     &   14.70   \\
66   &  MM1-II    &      -9.187  &   -435.588   &   2914.122     &   76.50   \\
67   &  MM1-II    &      -9.121  &   -431.400   &   2910.171     &   9.14    \\
68   &  MM1-II    &      -8.967  &   -670.744   &   3506.137     &   6.89    \\
69   &  MM3       &      -8.902  &    139.280   &   46.260       &   8.34    \\
70   &  MM3       &      -8.880  &    125.988   &  -11.458       &   7.50    \\
71   &  MM1-II    &      -8.660  &   -390.434   &   2860.458     &   14.40   \\
72   &  MM1-II    &      -8.565  &   -383.283   &   2854.336     &   13.40   \\
73   &  MM3       &      -8.506  &    734.114   &   -126.733     &   35.40   \\
74   &  MM1-II    &      -8.419  &   -578.443   &   3050.824     &   4.41    \\
75   &  MM1-II    &      -8.309  &   -402.934   &   2851.412     &   17.90   \\
76   &  MM3       &      -8.111  &    469.124   &  -1224.190     &   2.42    \\
77   &  MM3       &      -7.958  &    473.724   &  -1225.854     &   3.62    \\
78   &  MM1-I     &      -7.936  &    593.883   &   4401.716     &   4.78    \\
79   &  MM1-I     &      -7.914  &    609.700   &   4395.580     &   4.40    \\
80   &  MM1-I     &      -7.892  &    573.031   &   4132.954     &   2.82    \\
81   &  MM1-I     &      -7.892  &    355.203   &   4541.005     &   4.15    \\
82   &  MM1-I     &      -7.862  &    588.939   &   4393.666     &   5.26    \\
83   &  MM1-I     &      -7.847  &    597.584   &   4389.842     &   71.10   \\
84   &  MM1-I     &      -7.694  &    580.934   &   4397.245     &   5.53    \\
85   &  MM1-I     &      -7.474  &    560.840   &   4415.239     &   32.70   \\
86   &  MM1-I     &      -7.436  &    550.101   &   4410.280     &   93.60   \\
87   &  MM1-I     &      -7.343  &    534.182   &   4416.507     &   89.80   \\
88   &  MM1-I     &      -7.343  &    545.309   &   4421.374     &   70.60   \\
89   &  MM1-I     &      -7.340  &    564.765   &   4403.569     &   205.0   \\
90   &  MM1-I     &      -7.321  &    531.210   &   4427.860     &   34.20   \\
91   &  MM1-I     &      -7.296  &    551.174   &   4420.600     &   17.10   \\
92   &  MM1-I     &      -7.277  &    564.736   &   4412.451     &   13.10   \\
93   &  MM1-I     &      -7.205  &    554.632   &   4407.612     &   72.20   \\
94   &  MM1-I     &      -7.189  &    536.294   &   4424.718     &   19.00   \\
95   &  MM1-I     &      -7.145  &    555.577   &   4416.282     &   38.00   \\
96   &  MM1-I     &      -7.079  &    563.982   &   4407.374     &   22.20   \\
97   &  MM1-I     &      -7.057  &    314.574   &   4075.703     &   8.15    \\
98   &  MM1-I     &      -7.043  &    574.231   &   4409.231     &   13.60   \\
99   &  MM1-I     &      -6.969  &    579.531   &   4390.192     &   84.30   \\
100  &  MM1-I     &      -6.928  &    573.567   &   4398.639     &   62.60   \\
101  &  MM1-I     &      -6.878  &    590.635   &   4387.847     &   9.41    \\
102  &  MM1-I     &      -6.842  &    596.594   &   4383.680     &   46.90   \\
103  &  MM1-I     &      -6.838  &    582.891   &   4397.082     &   164.0   \\
104  &  MM1-I     &      -6.838  &    701.563   &   4427.330     &   7.13    \\
105  &  MM1-I     &      -6.829  &    332.840   &   4070.864     &   8.23    \\
106  &  MM1-I     &      -6.805  &    589.678   &   4395.311     &   20.70   \\
107  &  MM3       &      -6.772  &    603.937   &  -1980.915     &   9.55    \\
108  &  MM1-I     &      -6.762  &    596.730   &   4393.775     &   9.48    \\
109  &  MM3       &      -6.676  &    612.129   &  -1980.942     &   5.25    \\
110  &  MM3       &      -6.619  &    596.020   &  -1975.030     &   1.51    \\
111  &  MM3       &      -6.487  &    650.772   &  -2028.827     &   4.83    \\
112  &  MM1-III   &      -6.465  &    414.746   &   2742.622     &   13.00   \\
113  &  MM1-I     &      -6.448  &    783.752   &   5060.169     &   3.51    \\
114  &  MM1-III   &      -6.443  &    406.636   &   2750.507     &   4.19    \\
115  &  MM1-III   &      -6.281  &    388.405   &   2764.102     &   2.12    \\
116  &  MM1-III   &      -6.279  &    397.097   &   2758.756     &   4.26    \\
117  &  MM1-III   &      -6.201  &    377.350   &   2774.629     &   1.88    \\
118  &  MM1-III   &      -5.828  &    449.170   &   2515.400     &   3.68    \\
119  &  MM1-III   &      -5.762  &    466.783   &   2518.655     &   3.27    \\
120  &  MM1-III   &      -5.724  &    453.766   &   2515.285     &   49.90   \\
121  &  MM1-III   &      -5.699  &    456.314   &   2519.131     &   16.10   \\
122  &  MM1-I     &      -5.499  &    799.456   &   5047.388     &   2.47    \\
123  &  MM1-I     &      -5.323  &    937.690   &   4492.288     &   7.31    \\
124  &  MM1-I     &      -5.301  &    936.302   &   4485.963     &   5.04    \\
125  &  MM1-I     &      -5.096  &    879.986   &   4388.052     &   2.87    \\
126  &  MM1-I     &      -5.060  &    779.089   &   4540.142     &   6.20    \\
127  &  MM1-I     &      -4.994  &    779.440   &   4529.658     &   2.99    \\
128  &  MM1-I     &      -4.906  &    771.523   &   4537.757     &   4.60    \\
\hline
\bottomrule
\end{longtable}


\begin{longtable}{c|l|c|c|c|c}
\caption{Same as Table~\ref{tab:Detailsv255Iallmaserfeatures}, but in March 2020.}\label{tab:Detailsv255AIallmaserfeatures}\\
\toprule
\hline
\multicolumn{1}{c}{\bf Feature} &
\multicolumn{1}{c}{\bf feature}  &
\multicolumn{1}{c}{\bf Velocity}  &
\multicolumn{1}{c}{\bf R. A. Offset}   &
\multicolumn{1}{c}{\bf DEC. Offset}   &
\multicolumn{1}{c}{\bf Integrated Flux}  \\
\multicolumn{1}{c}{\bf Number} &
\multicolumn{1}{c}{\bf Association}  &
\multicolumn{1}{c}{\bf (kms$^{-1}$)}  &
\multicolumn{1}{c}{\bf (mas)}   &
\multicolumn{1}{c}{\bf (mas)}   &
\multicolumn{1}{c}{\bf (Jy~beam$^{-1}$)}  \\
\toprule
\hline
1  &    MM2      &  -11.447       &  -2344.689        &     2238.353        &  2.62   \\
2  &    MM2      &  -11.425       &  -2352.708        &     2241.711        &  15.70  \\
3  &    MM2      &  -11.398       &  -2369.039        &     2247.275        &  3.01   \\
4  &    MM2      &  -11.381       &  -2361.090        &     2244.261        &  14.00  \\
5  &    MM2      &  -11.373       &  -2217.252        &     2529.221        &  3.34   \\
6  &    MM2      &  -11.315       &  -2221.417        &     2531.696        &  5.51   \\
7  &    MM2      &  -11.271       &  -2387.757        &     2236.995        &  4.49   \\
8  &    MM2      &  -11.271       &  -2373.987        &     2229.949        &  3.33   \\
9  &    MM2      &  -11.249       &  -2359.823        &     2232.146        &  3.65   \\
10  &   MM2      &  -11.161       &  -2370.536        &     2235.519        &  8.07   \\
11  &   MM2      &  -11.161       &  -2119.187        &     2569.843        &  4.00   \\
12  &   MM2      &  -11.158       &  -2401.509        &     2237.835        &  27.00  \\
13  &   MM2      &  -11.141       &  -2378.819        &     2229.481        &  13.30  \\
14  &   MM2      &  -11.140       &  -2385.130        &     2247.859        &  4.96   \\
15  &   MM2      &  -11.139       &  -2158.504        &     2759.780        &  8.03   \\
16  &   MM2      &  -11.139       &  -2308.187        &     2292.005        &  9.85   \\
17  &   MM2      &  -11.124       &  -2366.418        &     2226.650        &  9.23   \\
18  &   MM2      &  -11.117       &  -2388.095        &     2232.843        &  6.12   \\
19  &   MM2      &  -11.091       &  -2370.622        &     2223.972        &  11.50  \\
20  &   MM2      &  -11.090       &  -2393.910        &     2232.045        &  54.60  \\
21  &   MM2      &  -11.089       &  -2383.985        &     2236.916        &  3.92   \\
22  &   MM2      &  -11.089       &  -2396.939        &     2240.177        &  6.64   \\
23  &   MM2      &  -11.074       &  -2539.549        &     1947.288        &  13.90  \\
24  &   MM2      &  -11.069       &  -2313.674        &     2293.120        &  11.90  \\
25  &   MM2      &  -11.066       &  -2150.387        &     2753.719        &  5.27   \\
26  &   MM2      &  -11.064       &  -2170.200        &     2763.201        &  8.58   \\
27  &   MM2      &  -11.051       &  -2304.601        &     2287.886        &  14.70  \\
28  &   MM2      &  -11.046       &  -2413.897        &     2242.075        &  15.00  \\
29  &   MM2      &  -10.986       &  -2406.028        &     2235.801        &  75.40  \\
30  &   MM2      &  -10.934       &  -2425.005        &     2246.428        &  3.63   \\
31  &   MM2      &  -10.788       &  -2371.960        &     2246.932        &  1.99   \\
32  &   MM2      &  -10.766       &  -2453.141        &     2231.668        &  3.47   \\
33  &   MM3      &  -10.744       &   110.646         &    -123.625         &  11.40  \\
34  &   MM3      &  -10.744       &  -73.237          &    -487.654         &  49.30  \\
35  &   MM3      &  -10.722       &  -80.250          &    -482.646         &  6.60   \\
36  &   MM3      &  -10.711       &  -46.984          &    -109.533         &  9.72   \\
37  &   MM2      &  -10.701       &  -2461.394        &     2236.770        &  4.31   \\
38  &   MM3      &  -10.679       &  -10.825          &    -320.187         &  10.60  \\
39  &   MM3      &  -10.671       &  -51.072          &    -106.898         &  71.30  \\
40  &   MM3      &  -10.657       &   106.576         &    -120.877         &  6.55   \\
41  &   MM3      &  -10.642       &   326.380         &     708.252         &  18.40  \\
42  &   MM3      &  -10.635       &  -55.485          &    -103.613         &  153.00 \\
43  &   MM3      &  -10.621       &   322.549         &     710.889         &  24.80  \\
44  &   MM3      &  -10.569       &  -484.716         &     213.634         &  14.50  \\
45  &   MM3      &  -10.569       &  -64.828          &    -93.188          &  13.50  \\
46  &   MM3      &  -10.547       &  -488.129         &     218.397         &  26.90  \\
47  &   MM3      &  -10.539       &  -306.938         &     14.627          &  6.12   \\
48  &   MM3      &  -10.481       &  -717.564         &     472.148         &  17.20  \\
49  &   MM3      &  -10.481       &   37.423          &    -43.655          &  37.00  \\
50  &   MM3      &  -10.459       &  -71.011          &    -247.669         &  29.60  \\
51  &   MM3      &  -10.459       &   254.890         &     493.152         &  21.30  \\
52  &   MM3      &  -10.459       &   234.850         &     502.496         &  18.70  \\
53  &   MM3      &  -10.437       &  -521.460         &     257.597         &  10.60  \\
54  &   MM3      &  -10.415       &  -2.067           &     165.218         &  13.30  \\
55  &   MM3      &  -10.393       &  -521.060         &     262.930         &  10.60  \\
56  &   MM3      &  -10.393       &  -1420.063        &     778.740         &  10.10  \\
57  &   MM3      &  -10.371       &   239.299         &     501.963         &  15.80  \\
58  &   MM3      &  -10.349       &  -104.376         &    -45.433          &  54.60  \\
59  &   MM3      &  -10.349       &  -95.973          &    -50.814          &  59.00  \\
60  &   MM3      &  -10.334       &   55.043          &    -215.503         &  15.50  \\
61  &   MM3      &  -10.331       &   233.880         &     510.149         &  16.30  \\
62  &   MM3      &  -10.326       &  -16.669          &     18.469          &  17.40  \\
63  &   MM3      &  -10.324       &  -30.046          &     19.373          &  17.00  \\
64  &   MM3      &  -10.291       &  -294.643         &    -589.471         &  16.00  \\
65  &   MM3      &  -10.284       &  -52.178          &    -253.794         &  20.90  \\
66  &   MM3      &  -10.271       &  -57.977          &    -248.493         &  20.50  \\
67  &   MM3      &  -10.262       &  -196.407         &    -89.533          &  30.10  \\
68  &   MM3      &  -10.262       &  -153.152         &    -294.860         &  13.10  \\
69  &   MM3      &  -10.261       &  -215.408         &    -531.070         &  18.60  \\
70  &   MM3      &  -10.261       &  -22.674          &     13.129          &  22.70  \\
71  &   MM3      &  -10.261       &   2.341           &    -1.879           &  73.40  \\
72  &   MM3      &  -10.239       &   7.672           &    -9.810           &  21.70  \\
73  &   MM3      &  -10.218       &  -76.779          &    -57.655          &  69.80  \\
74  &   MM3      &  -10.204       &  -2.672           &     6.019           &  186.00 \\
75  &   MM3      &  -10.203       &  -6.501           &     2.660           &  34.90  \\
76  &   MM3      &  -10.196       &   219.782         &     522.846         &  7.94   \\
77  &   MM3      &  -10.174       &  -1281.093        &     637.403         &  5.92   \\
78  &   MM3      &  -10.174       &  -511.780         &     262.835         &  6.01   \\
79  &   MM3      &  -10.136       &  -13.871          &     8.445           &  16.90  \\
80  &   MM3      &  -10.108       &  -4.448           &    -7.439           &  7.42   \\
81  &   MM2      &  -10.077       &  -2888.462        &     1479.054        &  5.09   \\
82  &   MM2      &  -10.077       &  -2884.855        &     1475.757        &  5.72   \\
83  &   MM3      &  -10.074       &  -114.330         &    -36.282          &  17.00  \\
84  &   MM3      &  -10.064       &  -17.823          &     15.714          &  21.00  \\
85  &   MM3      &  -10.006       &  -9.113           &     10.830          &  277.00 \\
86  &   MM2      &  -9.910        &  -2786.294        &     1364.479        &  2.45   \\
87  &   MM2      &  -9.888        &  -2391.033        &     2265.185        &  4.51   \\
88  &   MM3      &  -9.856        &  -12.187          &     13.145          &  9.45   \\
89  &   MM3      &  -9.808        &   83.000          &    -56.537          &  38.40  \\
90  &   MM3      &  -9.757        &  -108.768         &    -45.841          &  15.90  \\
91  &   MM3      &  -9.735        &   343.217         &     734.869         &  16.80  \\
92  &   MM3      &  -9.712        &   6.085           &    -6.344           &  46.50  \\
93  &   MM3      &  -9.703        &   224.932         &     533.310         &  22.80  \\
94  &   MM3      &  -9.655        &   244.131         &     524.915         &  10.20  \\
95  &   MM3      &  -9.652        &  -0.023           &    -0.207           &  445.00 \\
96  &   MM3      &  -9.625        &   108.756         &     45.618          &  13.10  \\
97  &   MM3      &  -9.515        &   164.498         &     50.499          &  74.20  \\
98  &   MM3      &  -9.515        &   171.570         &     45.505          &  10.20  \\
99  &   MM3      &  -9.493        &   157.410         &     57.318          &  8.45   \\
100  &  MM1-II   &  -9.449        &  -421.466         &     2583.902        &  4.34   \\
101  &  MM1-II   &  -9.383        &  -668.403         &     3514.066        &  23.80  \\
102  &  MM1-II   &  -9.375        &  -568.663         &     3566.226        &  4.19   \\
103  &  MM1-II   &  -9.340        &  -679.114         &     3520.492        &  4.28   \\
104  &  MM1-II   &  -9.010        &  -675.652         &     3506.023        &  15.20  \\
105  &  MM1-II   &  -8.966        &  -682.774         &     3509.845        &  8.28   \\
106  &  MM1-II   &  -8.923        &  -569.260         &     3650.489        &  1.98   \\
107  &  MM3      &  -8.901        &   138.826         &     46.282          &  6.41   \\
108  &  MM1-II   &  -8.873        &  -785.130         &     3300.372        &  4.72   \\
109  &  MM3      &  -8.857        &   125.472         &    -11.907          &  8.33   \\
110  &  MM1-II   &  -8.747        &  -655.326         &     3590.972        &  6.01   \\
111  &  MM1-II   &  -8.747        &  -646.604         &     3585.389        &  6.77   \\
112  &  MM3      &  -8.747        &   715.927         &    -106.807         &  5.00   \\
113  &  MM3      &  -8.725        &   710.330         &    -99.587          &  2.60   \\
114  &  MM1-II   &  -8.681        &  -639.071         &     3579.751        &  2.44   \\
115  &  MM3      &  -8.681        &   721.720         &    -115.081         &  3.53   \\
116  &  MM3      &  -8.593        &   728.121         &    -121.051         &  10.40  \\
117  &  MM3      &  -8.506        &   1085.203        &     604.111         &  5.51   \\
118  &  MM1-II   &  -8.506        &  -719.374         &     3546.517        &  3.56   \\
119  &  MM1-II   &  -8.471        &  -711.768         &     3540.953        &  6.35   \\
120  &  MM3      &  -8.471        &   734.848         &    -127.870         &  75.60  \\
121  &  MM3      &  -8.286        &   743.706         &    -131.573         &  7.24   \\
122  &  MM3      &  -8.176        &   468.360         &    -1224.892        &  4.96   \\
123  &  MM1-II   &  -8.176        &  -345.217         &     3672.169        &  1.33   \\
124  &  MM3      &  -7.957        &   757.299         &    -134.716         &  2.08   \\
125  &  MM3      &  -7.957        &   474.168         &    -1227.099        &  4.07   \\
126  &  MM3      &  -7.949        &   748.355         &    -130.218         &  3.13   \\
127  &  MM3      &  -7.858        &   752.229         &    -127.460         &  6.43   \\
128  &  MM1-I    &  -7.815        &   598.788         &     4389.730        &  19.30  \\
129  &  MM1-I    &  -7.815        &   605.960         &     4386.205        &  12.20  \\
130  &  MM1-I    &  -7.742        &   590.703         &     4394.947        &  7.47   \\
131  &  MM1-I    &  -7.737        &   589.740         &     4392.567        &  3.05   \\
132  &  MM1-I    &  -7.650        &   584.069         &     4397.790        &  3.90   \\
133  &  MM1-I    &  -7.532        &   419.175         &     4123.116        &  12.00  \\
134  &  MM1-I    &  -7.518        &   651.170         &     4463.942        &  9.64   \\
135  &  MM1-I    &  -7.515        &   636.686         &     4469.861        &  12.30  \\
136  &  MM1-I    &  -7.474        &   627.120         &     4473.235        &  8.91   \\
137  &  MM1-I    &  -7.474        &   508.829         &     4431.199        &  11.30  \\
138  &  MM1-I    &  -7.473        &   411.546         &     4126.178        &  16.50  \\
139  &  MM1-I    &  -7.452        &   394.552         &     4133.707        &  6.34   \\
140  &  MM1-I    &  -7.430        &   526.479         &     4422.340        &  22.90  \\
141  &  MM1-I    &  -7.424        &   643.462         &     4466.661        &  13.20  \\
142  &  MM1-I    &  -7.414        &   536.075         &     4418.142        &  202.00 \\
143  &  MM1-I    &  -7.401        &   546.506         &     4413.199        &  80.10  \\
144  &  MM1-I    &  -7.390        &   558.620         &     4407.557        &  32.90  \\
145  &  MM1-I    &  -7.364        &   517.880         &     4427.869        &  10.10  \\
146  &  MM3      &  -7.342        &   509.592         &     84.840          &  5.83   \\
147  &  MM1-I    &  -7.328        &   567.580         &     4402.631        &  28.00  \\
148  &  MM1-I    &  -7.320        &   319.490         &     4069.892        &  5.86   \\
149  &  MM3      &  -7.298        &   511.155         &    -1976.125        &  10.30  \\
150  &  MM1-I    &  -7.189        &   430.552         &     4114.260        &  5.71   \\
151  &  MM1-I    &  -7.189        &   421.557         &     4118.245        &  10.30  \\
152  &  MM1-I    &  -7.152        &   554.963         &     4406.019        &  40.20  \\
153  &  MM1-I    &  -7.108        &   313.910         &     4240.813        &  12.10  \\
154  &  MM1-I    &  -7.097        &   576.812         &     4397.547        &  3.50   \\
155  &  MM1-I    &  -7.079        &   559.180         &     4403.377        &  3.52   \\
156  &  MM1-I    &  -7.079        &   426.197         &     4539.023        &  5.83   \\
157  &  MM1-I    &  -7.079        &   415.666         &     4542.187        &  7.11   \\
158  &  MM1-I    &  -7.057        &   553.161         &     4311.414        &  18.50  \\
159  &  MM1-I    &  -7.057        &   542.294         &     4317.776        &  16.70  \\
160  &  MM1-I    &  -7.035        &   436.614         &     4535.426        &  6.95   \\
161  &  MM1-I    &  -7.013        &   576.580         &     4394.810        &  10.00  \\
162  &  MM1-I    &  -7.013        &   572.920         &     4397.565        &  15.60  \\
163  &  MM1-I    &  -7.004        &   321.332         &     4237.441        &  7.14   \\
164  &  MM1-I    &  -6.976        &   461.346         &     4525.363        &  25.80  \\
165  &  MM1-I    &  -6.969        &   433.192         &     4441.777        &  29.90  \\
166  &  MM1-I    &  -6.969        &   583.600         &     4391.375        &  9.93   \\
167  &  MM1-I    &  -6.947        &   579.900         &     4393.540        &  17.50  \\
168  &  MM1-I    &  -6.934        &   332.674         &     4232.878        &  6.35   \\
169  &  MM1-I    &  -6.925        &   425.251         &     4447.496        &  17.60  \\
170  &  MM3      &  -6.896        &   604.470         &    -1982.235        &  7.40   \\
171  &  MM1-I    &  -6.881        &   444.343         &     4440.041        &  4.13   \\
172  &  MM1-I    &  -6.859        &   608.110         &     4379.739        &  5.86   \\
173  &  MM1-I    &  -6.845        &   588.394         &     4388.932        &  34.70  \\
174  &  MM1-I    &  -6.838        &   457.412         &     4097.404        &  6.31   \\
175  &  MM1-II   &  -6.838        &  -359.031         &     2493.480        &  3.59   \\
176  &  MM3      &  -6.811        &   613.429         &    -1985.703        &  15.80  \\
177  &  MM1-I    &  -6.801        &   596.478         &     4382.635        &  44.30  \\
178  &  MM1-I    &  -6.794        &   466.816         &     4523.512        &  17.50  \\
179  &  MM1-II   &  -6.772        &  -336.567         &     2577.340        &  7.45   \\
180  &  MM3      &  -6.706        &   619.730         &    -1990.288        &  6.14   \\
181  &  MM1-I    &  -6.662        &   613.463         &     4371.253        &  4.97   \\
182  &  MM1-I    &  -6.662        &   372.442         &     4312.386        &  5.55   \\
183  &  MM3      &  -6.618        &   650.144         &    -2025.959        &  9.66   \\
184  &  MM1-I    &  -6.618        &   340.079         &     4510.960        &  8.48   \\
185  &  MM1-I    &  -6.618        &   332.647         &     4515.790        &  4.22   \\
186  &  MM1-I    &  -6.596        &   358.252         &     4505.727        &  7.55   \\
187  &  MM3      &  -6.596        &   871.362         &    -1948.936        &  7.84   \\
188  &  MM1-I    &  -6.552        &   346.808         &     4508.783        &  8.55   \\
189  &  MM3      &  -6.549        &   890.405         &    -1862.921        &  6.90   \\
190  &  MM1-I    &  -6.472        &   333.769         &     4502.903        &  3.20   \\
191  &  MM1-I    &  -6.428        &   347.083         &     4489.780        &  3.45   \\
192  &  MM1-I    &  -6.399        &   363.685         &     4486.787        &  2.80   \\
193  &  MM1-I    &  -6.398        &   328.461         &     4497.470        &  26.00  \\
194  &  MM1-I    &  -6.389        &   321.010         &     4501.525        &  7.43   \\
195  &  MM1-I    &  -6.369        &   314.426         &     4509.004        &  3.24   \\
196  &  MM1-I    &  -6.340        &   313.994         &     4493.140        &  6.68   \\
197  &  MM1-I    &  -6.333        &   341.575         &     4495.285        &  8.90   \\
198  &  MM1-I    &  -6.333        &   320.703         &     4486.623        &  11.50  \\
199  &  MM1-I    &  -6.267        &   176.729         &     4199.946        &  2.85   \\
200  &  MM1-I    &  -6.223        &   300.967         &     4493.294        &  14.00  \\
201  &  MM1-I    &  -6.223        &   409.554         &     4540.088        &  5.31   \\
202  &  MM1-I    &  -6.223        &   186.833         &     4193.138        &  2.62   \\
203  &  MM1-I    &  -6.190        &   319.809         &     4482.615        &  4.95   \\
204  &  MM1-I    &  -6.168        &   312.718         &     4486.292        &  15.70  \\
205  &  MM1-I    &  -6.128        &   307.728         &     4488.494        &  23.30  \\
206  &  MM1-I    &  -5.015        &   771.441         &     4537.619        &  11.80  \\
\hline
\bottomrule
\end{longtable}

\twocolumn


\bsp	
\label{lastpage}
\end{document}